\documentstyle[preprint,aps]{revtex}

\begin{document}

\draft

\preprint{SI-94-TP3S2; STPHY-Th/94-8}

\title   {Loosely bound hyperons in the SU(3) Skyrme model}

\author{
B. Schwesinger$^{1,2}$\footnote{Electronic address:
schwesinger@hrz.uni-siegen.d400.de},
F.G. Scholtz$^1$\footnote{Electronic address:
fgs@sunvax.sun.ac.za}
and
H.B. Geyer$^1$\footnote{Electronic address:
hbg@sunvax.sun.ac.za}
}

\address {               $^1$Institute of Theoretical Physics,
                            University of Stellenbosch,
                           Stellenbosch 7600, South Africa
}
\address{               $^2$Siegen University, Fachbereich Physik,
                                  57068 Siegen, Germany }

\maketitle

\begin{abstract}
Hyperon pairs bound in deuteron like states are obtained within
the SU(3) Skyrme model in agreement with general expectations from
boson exchange models. The central binding from the
flavor symmetry breaking terms increases with the strangeness contents
of the interacting baryons whereas the kinetic non-linear
$\sigma$-model term fixes the spin and
isospin of the bound pair. We give a complete account of the
interactions of octet baryons within the product approximation to
baryon number $B=2$ configurations.
\vspace{2cm}
\end{abstract}


\narrowtext

\section{Introduction}
\label{sec1}

It seems to be generally recognized that the relative scarcity of
experimental data on hyperon-nucleon and especially hyperon-hyperon
interactions calls for major theoretical input to arrive at a
satisfactory description of the data\cite{Nijm92}.  From the boson
exchange point of view such a description is typically approached by
``upgrading" a reliable description of the $N\!N$-interaction with
SU(3) flavor symmetry, assuming of course that the formulation of the
$N\!N$-interaction admits such a generalization.  An example of this
approach can be found in the construction of the Nijmegen potentials
(see Ref.\cite{Nijm92} and references therein).  The one-boson
exchange potentials obtained in this way generally predict the
existence of loosely bound hyperons, while ruling out deeply bound
hyperons.

An alternative approach which is well tailored to address the
construction of hyperon-nucleon and hyperon-hyperon potentials is one
based on the Skyrme model\cite{HS86,MZ86,Russians}.  It has already
been successfully implemented in the construction of the
nucleon-nucleon potential, as reviewed recently by Walhout and
Wambach\cite{WW92}.  In the Skyrme model deeply bound hyperons also
seem to be excluded once Casimir energies are properly taken into
account\cite{SSG93}, but deuteron-like states, for which a Skyrmion
product ansatz description seems to suffice, remain possible.  An
analysis of such configurations to determine hyperon-nucleon and
hyperon-hyperon potentials have in fact recently been carried out by
K\"albermann and Eisenberg\cite{KE90,KE92} where much of the focus is
on the central interactions.  No conclusive statements about the
existence or otherwise of bound states seem to have emerged from these
two studies.

Before we indicate below that the SU(3) Skyrme model construction of
hyperon-nucleon and hyperon-hyperon potentials can indeed lead to
qualitative and quantitative results, it is perhaps worthwhile to
point out that such an approach has the further merit that it ties
together the mesonic sector ($B=0$), the baryonic sector ($B=1$) and
the $B=2$ sector in the single comprehensive framework of a
dynamically closed model with only a few (typically~5) parameters.
Furthermore the SU(3) flavor symmetry breaking part is fixed by meson
masses and weak decay constants, while in turn it then reproduces the
hyperon spectrum in the $B=1$ sector and will also be responsible for
two hyperons forming loosely bound deuteron-like states.

We now turn to some of the more salient aspects of our analysis.  It
is rather satisfactory to find that, apart from the known problems
associated with the central attraction in the Skyrme model (see Sec.\
\ref{sec4} for further discussion of this point), many of the more
robust predictions from the one-boson exchange approach to the problem
also emerge from the Skyrme model analysis.  In this regard we mention
that we extract unambiguously an attractive $\Lambda\Lambda$
interaction in the $^1S_0$ state, a result also required from a number
of complementary analyses\cite{DMG+91a}.  Secondly we find e.g. an
$N\Sigma$ potential with relative components of the various spin and
isospin interactions which are comparable with those found in the
Nijmegen potentials\cite{DG84,DMG+91b}.  As another specific example,
the spin dependence of the $N\!\Lambda$ potential we obtain is the same
as that favoured from phenomenological and potential model estimates
in Ref.\cite{MGD+85}.  Furthermore it is possible to predict from our
analysis the spin and isospin of bound deuteron type hyperon states.
We also extract a rather prominent correlation of an increased binding
of baryons with growing strangeness.

A short outline of the presentation is as follows.  In Secs.
\ref{sec2} and \ref{sec3} we introduce the SU(3) lagrangian, recall
aspects of the framework in which it is typically employed and discuss
our strategy of identifying a set of independent relative orientations
of two skyrmions which eventually facilitates the identification of
the various interaction potentials.  This strategy, which generalises
the known approach for the $N\!N$-potential\cite{WW92,VM85}, seems
to offer advantages over the approach of Refs.\cite{KE90,KE92} where a
Monte Carlo calculation is used to evaluate matrix
elements\cite{Kal94}.  (See also Sec.\  \ref{sec4}.)  Detail of this
rather technical part is collected in Appendix ~\ref{secappA}.  In Sec.\
\ref{sec4} we consider the question of missing intermediate range
attraction in the Skyrme model and possible resolutions, argue to
which extent various aspects of our present analysis of
hyperon-hyperon interactions would retain their validity should this
be resolved satisfactorily, and then discuss general and specific
results for the various interactions.  Tables for the various
interaction potentials as function of the separation $R$ are collected
in Appendix ~\ref{secappB}.

\section{Lagrangian and $B=1$ sector}
\label{sec2}
The lagrangian we use is the standard SU(3) Skyrme-model lagrangian
given by\cite{JJPSW89,PSW91}
\begin{equation}\label{su3lag}
{\cal L} =
{\cal L}^{({\rm SYM})} +
{\cal L}^{({\rm FSB})}
\end{equation}
In terms of the chiral field $U$ and with the standard notation for
the left current
\begin{equation}
L_{\mu} = U^\dagger \partial_{\mu} U = i \lambda_a L_{\mu}^a\; ,
\end{equation}
and the topological current
\begin{equation}
B^{\mu} = \frac{\varepsilon^{\mu \nu \alpha \beta}}{24 \pi^2} \;{\rm
tr}\; L_\nu L_{\alpha} L_{\beta}\; ,
\end{equation}
the flavor symmetric and flavor symmetry breaking parts are given by
\begin{eqnarray}\label{lsym}
{\cal L}^{({\rm SYM})} (U) &=&
- \frac{f^2_\pi}{4} \;{\rm tr}\; L_{\mu} L^{\mu}
+\frac{1-\chi}{32e^2} \;{\rm tr}\; [L_{\mu}, L_{\nu}][L^{\mu} , L^\nu]
\nonumber\\
& &
+ \frac{\chi}{16e^2} \left\{ (\;{\rm tr}\; L_{\mu} L^{\mu})^2 - \;{\rm tr}\;
L_{\mu}L_{\nu}
\cdot \;{\rm tr}\; L^{\mu} L^\nu  \right\}
\nonumber\\
& &
- \frac{\varepsilon_6^2}{2} B_{\mu} B^{\mu}
+ \frac{m_{\pi}^2 f_{\pi}^2}{4} \;{\rm tr}\;( U + U^\dagger - 2)
\nonumber\\
&\equiv&{\cal L}^{({2})}+{\cal L}^{({\rm 4A})}+{\cal L}^{({\widetilde{\rm
4A}})}+{\cal L}^{({6})}+{\cal L}^{({\rm CSB})}
\end{eqnarray}
and
\begin{eqnarray}\label{lfsb}
{\cal L}^{({\rm FSB})} (U) & = &
- \frac{f_K^2 - f_{\pi}^2}{4} \;{\rm tr}\;
{
\renewcommand{\arraystretch}{.2}
\renewcommand{\arraycolsep}{2 pt}
\mbox{$
\left(
 \begin{array}{ccc}
  {\scriptstyle 0} & {\scriptstyle } & {\scriptstyle } \\
  {\scriptstyle } & {\scriptstyle 0} & {\scriptstyle } \\
  {\scriptstyle } & {\scriptstyle } & {\scriptstyle 1}
 \end{array}
\right)$}}
( U L_{\mu} L^{\mu} + L_{\mu} L^{\mu} U^\dagger)\nonumber\\
&&+ \frac{f_K^2 m_K^2  - f_{\pi}^2 m_{\pi}^2}{2} \;{\rm tr}\;
{
\renewcommand{\arraystretch}{.2}
\renewcommand{\arraycolsep}{2 pt}
\mbox{$
\left(
 \begin{array}{ccc}
  {\scriptstyle 0} & {\scriptstyle } & {\scriptstyle } \\
  {\scriptstyle } & {\scriptstyle 0} & {\scriptstyle } \\
  {\scriptstyle } & {\scriptstyle } & {\scriptstyle 1}
 \end{array}
\right)$}}
( U + U^\dagger - 2),
\end{eqnarray}
where we have provided for the possibility of different stabilizing
terms at most quadratic in the time derivatives.
Among the stabilizers of fourth order the extension to SU(3) allows
for an alternate term ${\cal L}^{({\widetilde{\rm
4A}})}$\cite{GL85} which reduces to the Skyrme term
${\cal L}^{({\rm 4A})}$ in the SU(2)-limit. It's effects in the $B=1$
sector\cite{PSW91} have been small, whereas here they are
shown to be large if $\chi$ is chosen accordingly.

The lagrangian (\ref{su3lag}) succesfully describes the baryon
splittings in the $B=1$
sector when empirical meson masses and decay constants are used.
Fig.\  \ref{fig1} shows the baryon spectrum calculated by treating the
flavor symmetry breaking in two different ways: for rigidly rotated
solitons (RRA)\cite{YA88,WSP90,PSW91} the chiral angle
minimizing the soliton in the
isospin subgroup is rotated without changes into flavor directions of
decreasing hypercharge. In case of the
slow rotator approach (SRA)\cite{SRA1,SRA2} the soliton is allowed to
change its shape according to the flavor symmetry breaking forces
excerted at a given strangeness angle. These forces mainly reduce the
one pion tail of the SU(2) soliton to a more rapid decay
corresponding to the higher meson masses present in the cloud.

\section{Product ansatz, energy functional and
asymptotic interaction in the $B=2$ sector}
\label{sec3}

In order to obtain a hyperon-hyperon potential (generally
baryon-baryon potential) we follow the standard Skyrme model
procedure, namely evaluate the energy of a $B=2$ configuration and
subtract the energies of the two $B=1$ configurations which constitute
that $B=2$ configuration. As the extremely short distance
behaviour of the potentials to be calculated is not of immediate
concern,
we content ourselves with the following hedgehog product ansatz for
the $B=2$ configuration

\begin{equation}
U ({\bbox{r}};A,C) = A U_{\rm H} ( x, y, z - \case{R}{2})
C \, U_{\rm H}       ( x, y, z + \case{R}{2})
C^\dagger A^\dagger
\end{equation}
where $U_{\rm H}(\bbox{r}) =  e^{i \bbox{\tau}\cdot \hat{r} \chi (r)}$. $A$
and $C=A^\dagger B$ are constant flavor rotations, $C$ denoting the
relative orientation of the two SU(3) skyrmions in flavor space.  As
a matter of convenience we chose the two skyrmions to be separated by
a distance $R$ along the $z$-axis.  The profile function for the
hedgehog is obtained by a numerical minimization of the energy
functional in the $B=1$ sector.  Using the product ansatz we can
calculate the energy of the $B=2$ configuration for any given
orientations $A$, $C$ and separation $R$ by performing a numerical
integration over the spatial coordinates of the hedgehog
configuration.

One notes from translation and flavor symmetry that
the energy functional resulting from ${\cal L}^{({\rm SYM})}$ depends only
on the relative separation $R$ and orientation $C$ of the two
skyrmions.  The
general structure of the energy functional is therefore established in
the form

\widetext

\begin{eqnarray}
V^{({\rm SYM})} (R;A,C)
&=&
U^{(0)} (R) + U_S^{(1)}(R) D_{ii}(C) + \Bigl( 3 D_{33} (C) -
D_{ii}(C)\Bigr) U_T^{(1)}(R)\nonumber\\
& & + \sum_{(I) = (4A), (\widetilde{4A}), (6)} \Bigl\{
U_{S}^{(2,I)} (R) \,
Q^{(I)} \case{1}{9} D_{ij}(C)D_{ij}(C) \nonumber\\
& &
\qquad +U_{S'}^{(2,I)} (R) \,
Q^{(I)} \case{1}{4} \Bigl(D_{ii}(C)D_{jj}(C) -D_{ij} (C) D_{ji} (C)
  \Bigr) \nonumber\\
& & \qquad + U_{T}^{(2,I)} (R) \,
Q^{(I)} \case{1}{4} \Bigl(6D_{11}(C)D_{22}(C) - 6 D_{12} (C) D_{21}
 (C)\nonumber\\
& & \qquad -D_{ii} (C) D_{jj} (C) + D_{ij} (C) D_{ji} (C) \Bigr)
  \Bigr\}\\
& & +U_{S}^{(3)} (R) D_{88} (C)\quad .\nonumber
\end{eqnarray}
Here we have used the convention $(i,j) \in \{1,2,3\}$
and the notation
\begin{equation}
Q^{(I)} = \left\{
\begin{array}{cl}
3 - \case{1}{2}L_a^2\quad ,  & (I) = (4A)\\
1\quad ,                     & (I) = ( \widetilde{4A}),\; (6)
\end{array}
\right.
\end{equation}
where $L_a^2$ is the Casimir operator of SU(3)
and $D_{ab}$ the SU(3) $D-$functions in the adjoint representation.
The occurence of the Casimir operator of SU(3) is due to a
manipulation
which allows us to keep the indices of the $D$-functions in the
SU(2)-subgroup: the energy density from the Skyrme term
${\cal L}^{({\rm 4A})}$ has
contributions proportional to
\begin{eqnarray}
f_{abc} D_{cc'} f_{ade} D_{ee'} &=&
- \Bigl( L_a D_{bc'} \Bigr)\Bigl( L_a D_{de'} \Bigr)\nonumber\\
&=& -\case{1}{2} L_a^2 \Bigl( D_{bc'} D_{de'}\Bigr)
+\case{1}{2} \Bigl( L_a^2 D_{bc'} \Bigr) D_{de'}
+ \case{1}{2}  D_{bc'} \Bigl(L_a^2D_{de'}\Bigr) \; ,
\end{eqnarray}
where $L_a$ are the left Euler angular momentum operators. The explicit
expressions stemming from the action of the Casimir operators are given
in Appendix \ref{secappA}.

\narrowtext

In Eq.\ (7) we
do not show those terms which contain combinations of
$D-$functions which are of second
rank in the spin in the energy functional, since their matrix elements
between the spin one-half baryon states vanish.  In principle they do,
however,
occur and due care has to be exercised, when the different components
of the
interaction are evaluated.  For completeness we list these combinations
ordered according to the spin-tensorial decomposition of the left and
right indices of the two $D$-functions which refer to the Euler angles
in $A$ and $B$ respectively.

\widetext

\begin{eqnarray}\label{dfuncs}
\Bigl[0 \times 2\Bigr]_2\;:\qquad&&D_{i3} D_{i3}-\case{1}{3}D_{ij} D_{ij}\;
,\nonumber\\
\Bigl[2 \times 0\Bigr]_2\;:\qquad&&D_{3i} D_{3i}-\case{1}{3}D_{ij} D_{ij}\;
,\nonumber\\
\Bigl[2 \times 2\Bigr]_0\;:\qquad
&&\case{1}{2}D_{ii} D_{jj}+\case{1}{2}D_{ij} D_{ji}-\case{1}{3}D_{ij} D_{ij}
\; ,\nonumber\\
\Bigl[2 \times 2\Bigr]_2\;:\qquad
&&\case{1}{2}D_{ii} D_{33}+\case{1}{2}D_{i3} D_{3i}-\case{1}{3}D_{i3} D_{i3}
-\case{1}{3}D_{3i} D_{3i}+\case{1}{9}D_{ij} D_{ij}\; ,\nonumber\\
\Bigl[2 \times 2\Bigr]_4\;:\qquad
&&D_{33} D_{33}-\case{1}{3}D_{i3} D_{i3}-\case{1}{3}D_{3i} D_{3i}
+\case{1}{9}D_{ij} D_{ij}\; .
\end{eqnarray}

The seven different components $U^{(0)}(R),U^{(1)}_S(R)\ldots U^{(2)}_T(R)$ of
the interaction appearing in $V^{({\rm SYM})} (R;A,C)$ can now be
calculated.
We note that if we evaluate the energy of the $B=2$ hedgehog-hedgehog
configuration for twelve orientations, $C$, such that the seven combinations of
$D-$functions appearing in $V^{({\rm SYM})} (R;A,C)$ and the five
combinations of
Eq.\ (\ref{dfuncs}) are linearly independent, we obtain a system of
linear equations
from which the interaction components can be solved by inversion of the
coefficient matrix.   The set of orientations
\begin{eqnarray}
&& 1\; ,\qquad e^{i \frac{\pi}{2} \lambda_2}\; ,\qquad e^{i
   \frac{\pi}{2} \lambda_3} \; ,\qquad e^{i \frac{\pi}{2} \lambda_4}
   e^{i \frac{\pi}{4} \lambda_2}\; ,\quad
   e^{i \frac{2\pi}{3} \lambda_2}e^{i \frac{\pi}{2}
   \lambda_4}
   \; ,\qquad e^{i \frac{\pi}{4} \lambda_4}e^{i \frac{\pi}{2}
   \lambda_2} e^{-i \frac{\pi}{4} \lambda_4}\; ,\nonumber\\
&& e^{i \frac{\pi}{4} \lambda_4}e^{i \frac{\pi}{2} \lambda_3}
   e^{-i \frac{\pi}{4} \lambda_4}
   \; ,\qquad e^{i \frac{\pi}{4} \lambda_1}e^{i \frac{\pi}{4}
   \lambda_4} e^{i \frac{\pi}{4} \lambda_2}\; ,\qquad
   e^{i \frac{\pi}{6} \lambda_4}e^{i \frac{\pi}{5} \lambda_6}
   e^{i \frac{\pi}{5} \lambda_4}
   \; ,\qquad e^{i \frac{\pi}{6} \lambda_1}e^{i \frac{\pi}{5}
   \lambda_6} e^{i \frac{\pi}{5} \lambda_4}\; , \nonumber\\
&& e^{i \frac{\pi}{6} \lambda_2}e^{i \frac{\pi}{5}
   \lambda_6} e^{i \frac{\pi}{5} \lambda_4}
   \; ,\qquad e^{i \frac{\pi}{6} \lambda_3}e^{i \frac{\pi}{5}
   \lambda_6} e^{i \frac{\pi}{3} \lambda_4}\; .
\end{eqnarray}
can be checked to give the required linear independence and are
subsequently used to determine the interaction components as already
outlined.

Translation invariance implies that for the flavor symmetry breaking
part ${\cal L}^{({\rm FSB})}$, the energy functional depends on $R$
alone, but since flavor symmetry is broken, it depends on the
orientations $A$ and $B$ of the individual skyrmions in flavor space.
Neglecting the numerically small and cumbersome terms proportional to
$f_K^2-f_\pi^2$, a simple calculation shows the energy functional to
be
\begin{eqnarray}
V^{({\rm FSB})} (R;A,B)
&=&
   U_0^{({\rm FSB})} (R)
   \Bigl[ \Bigl(1 - D_{88} (A)\Bigr)
   \Bigl(1 - D_{88} (B)\Bigr) + \case{3}{4} D_{k8} (A) D_{k8} (B)
   \Bigr] \nonumber \\
&+&
   U_1^{({\rm FSB})} (R)
   \Bigr[ D_{8i} (A) D_{8i} (B)+\case{3}{4} D_{ki} (A) D_{ki} (B)
   \Bigr] \nonumber\\
&+&
   U_2^{({\rm FSB})} (R) \Bigl[
   \Bigl( 3 D_{83} (A) D_{83} (B) - D_{8i} (A) D_{8i} (B)\Bigr)
\nonumber \\
& &\phantom{U_2^{({\rm FSB})} (R) \Bigl[ }
+\case{3}{4} \Bigl( 3 D_{k3} (A) D_{k3} (B) - D_{ki} (A) D_{ki}
(B)\Bigr) \Bigr]\; ,
\end{eqnarray}
where $k \in \{4,5,6,7\},\, i \in \{1,2,3\}$.
The radial functions in the flavor symmetry breaking part of
the energy functional, $V^{({\rm FSB})} (R;A,B)$, are given in terms
of the integrals
\begin{eqnarray}
U_0^{({\rm FSB})} (R)&=&\frac{4}{9}\Bigl( f_K^2 m_K^2 -f_\pi^2 m_\pi^2 \Bigr)
\int -(1-c_1)(1-c_2)\,d^3r \\
U_1^{({\rm FSB})} (R)&=&\frac{4}{9}\Bigl( f_K^2 m_K^2 -f_\pi^2 m_\pi^2 \Bigr)
\int \hat r(1)\cdot \hat r(2) s_1 s_2 \, d^3r \nonumber \\
U_2^{({\rm FSB})} (R)&=&\frac{4}{9}\Bigl( f_K^2 m_K^2 -f_\pi^2 m_\pi^2 \Bigr)
\int \case{1}{2}\Bigl(3\hat r_3(1) \hat r_3(2) -\hat r(1) \cdot \hat r(2)\Bigr)
s_1 s_2 \, d^3r \nonumber \; ,  \\
& & {\bbox r}(i) = {\bbox r}\pm\frac{R}{2} {\bbox e}_3,\quad
c_i= \cos \chi\bigl(r(i)\bigr),\quad s_i= \sin \chi\bigl(r(i)\bigr)
\quad i=1,2 \nonumber \; .
\end{eqnarray}
These integrals are evaluated numerically.

\narrowtext

Finally, knowing the interaction components, we have to evaluate the matrix
elements of the energy functional between the spin one-half baryon
wavefunctions diagonalizing the rotational
hamiltonian with the symmetry breaking term, see \cite{YA88,WSP90,PSW91}
and Eq.\ (A1) to
extract the potentials in the various interaction channels.
We have to distinguish direct and
exchange matrix elements, see Fig.\  \ref{fig2}. The single particle
degrees of
freedom of the interacting baryons are given by the collective rotations
$A$ and $B$ respectively,
which appear in all the $D$-functions of the potential energy.
In such an interaction term $D(A)$, say, can transfer strangeness
$0 \leq \vert S \vert \leq 1$ to the incoming baryon state depending on
the Euler angles $A$. For
the direct terms the outgoing baryon also carrying the Euler angles $A$ has
the same hypercharge as the incoming one, whereas the exchange terms
allow for different strangeness of the outgoing baryon.
Since total strangeness is, of course, conserved, the baryon
described by the Euler angles $B$ must compensate for the hypercharge
difference. Speaking in terms of a one boson exchange description, the
latter processes thus contain the kaon exchange terms.

The details of the calculation of the
matrix elements and the various interaction potentials are collected
in Appendix  \ref{secappA}. We must emphasize,
however, that generally, we will only use wave functions
diagonalizing the rigid rotator hamiltonian, because the use of a
chiral angle which changes in response to the flavor
symmetry breaking forces\cite{SRA1,SRA2} is technically too complicated, here.
For testing purposes we will also use
baryon wavefunctions which are SU(3) symmetric as well as those
for the strong symmetry breaking limit.
Unfortunately, the restriction to the rigidly rotating soliton
directly leads to the fact that the
asymptotic forces will always have the one-pion exchange tail, even for
kaon exchange terms.
This statement
can be verified by examination of the asymptotic interaction, which
is related to the lower order terms in the lagrangian after insertion of the
equations of motion for the individual solitons. Within SU(2) this
has been shown by Yabu and Ando \cite{YA85}, and the extension to
SU(3) is straight-forward:

\widetext

\begin{eqnarray}
&&V(R;A,B) \stackrel{R\rightarrow \infty}{\longrightarrow} \\
&&\qquad-\case{1}{2} \int {\rm tr} \; \Bigl(
f_\pi^2 \nabla \lambda_a \Phi_a(1) \cdot \nabla \lambda_b \Phi_b(2)
+f_\pi^2
{
\renewcommand{\arraystretch}{.2}
\renewcommand{\arraycolsep}{2 pt}
\mbox{$
\left(
 \begin{array}{ccc}
  {\scriptstyle m_\pi^2} & {\scriptstyle } & {\scriptstyle } \\
  {\scriptstyle } & {\scriptstyle m_\pi^2} & {\scriptstyle } \\
  {\scriptstyle } & {\scriptstyle } & {\scriptstyle 2 m_K^2-m_\pi^2}
 \end{array}
\right)$}}
\case{1}{2} \{ \lambda_a \Phi_a(1),\lambda_b \Phi_b(2) \} \Bigr) \; d^3r
\; .\nonumber
\end{eqnarray}
The eight asymptotic pseudoscalar soliton fields $\Phi_a$
solving the
Euler Lagrange equations should have the form\footnote{We thank A. Hayashi
for supplying us the asymptotic solution of the Euler Lagrange
equations when flavor symmetry is broken.}
\begin{eqnarray}
\Phi_a = D_{ab}\hat r_b \left\{
\begin{array}{ll}
 \chi^{\phantom K}_\pi & a\in \{1,2,3\}\\
 \chi^{\phantom K}_K & a\in \{4,5,6,7\}\\
 \chi^{\phantom K}_\eta & a=8
\end{array}
\right.
\end{eqnarray}
with
\begin{eqnarray}
\hat r \chi^{\phantom K}_C = -\nabla A_C \frac{{\rm e}^{-m_Cr}}{r}
\end{eqnarray}
leading to
\begin{eqnarray}
&&V(R;A,B) \stackrel{R\rightarrow \infty}{\longrightarrow} \\
&&\qquad 4 \pi f_\pi^2 \nabla_i \nabla_j \left\{
D_{ki}(A)D_{kj}(B) A_K^2 \frac{{\rm e}^{-m_K R}}{R} \right. \nonumber\\
&& \qquad \qquad \qquad \qquad \left.
+D_{8i}(A)D_{8j}(B) A_\eta^2 \frac{{\rm e}^{-m_\eta R}}{R}
+D_{i'i}(A)D_{i'j}(B) A_\pi^2 \frac{{\rm e}^{-m_\pi R}}{R}
\right\} \; . \nonumber
\end{eqnarray}
However, due to the rigid rotator approximation all masses are set to
$m_\pi$ and the asymptotic constants all equal
\begin{equation}
A_{\rm SU(2)}= \frac{g_{\pi NN}(-m_\pi^2)}{g_{\pi NN}(0)}
\; \frac{3 g_A^{\rm SU(2)}}{8 \pi f_\pi^2} \; .
\end{equation}
For SU(3)-symmetry the matrix elements of the $D$-functions in Eq.\ (17)
for nucleons are summarized by
\begin{eqnarray}
D_{ij}(C)=D_{ai}(A)D_{aj}(B) \rightarrow
(\case{7}{30})^2 \sigma_i(A) \sigma_j(B) \Bigl[
{\bbox \tau}(A) \cdot {\bbox \tau}(B)
+\case{3}{49}\Bigr]
\end{eqnarray}
and we recover the conventional one pion exchange interaction together
with a small term from one $\eta$ exchange  since for
SU(3)-symmetry $g_A^{\rm SU(3)}=\case{7}{10} g_A^{\rm SU(2)}$ in the
Skyrme model.

Numerically and fortunately,
the long range part of the exchange matrix elements will turn out to be
small, so the impact of the wrong tail is not important.

\narrowtext

\section{Results}
\label{sec4}
The main emphasis of the present investigation is on the question
of whether strangeness is loosely bound. Accessing the problem within
the framework of the Skyrme model has some advantageous and some
problematic aspects.

The problematic aspects stem from uncertainties in the higher
order terms of the lagrangian, inadequacies of the product ansatz
and, ultimately, the disturbing feature that an intermediate range
attraction in the nucleon-nucleon system does not emerge as simply and
naturally as one might have hoped , see e.g.
\cite{WW92,VM85,JJP85,YSH89} and references therein.
The obvious advantages lie in the fact that the lower order terms of
the lagrangian are unambiguously fixed in the mesonic winding number
zero sector.
Now, the same lagrangian governs the higher winding number configurations,
so that conclusions based solely on the lower order
terms, should also be reliable for the case of higher winding numbers.
One example for this is the baryon spectrum
in SU(3), where the lower order SU(3) symmetry breaking in the
lagrangian, Eq.\ (\ref{lfsb}), is sufficient to describe the mass
splittings,
as recalled in Sec.\ \ref{sec2}. Another example is given in the case of
meson-baryon scattering where
the use of mesonic weak decay constants and masses for this winding
number one configuration is mandatory\cite{Wal91,S92}.

Provided that the problem of the missing intermediate range attraction
finds its resolution in higher order terms and/or in improved
two-baryon configurations\cite{WW92} beyond the product ansatz, we
can confidently focus on the longer range part of the interaction:
it does not involve the higher order terms explicitely and also the
product ansatz gradually becomes exact.

The longer range components of the baryon-baryon forces are due to the
flavor symmetric and flavor symmetry breaking
mass terms ${\cal L}^{({\rm CSB})}$ and ${\cal L}^{({\rm FSB})}$,
and the non-linear $\sigma$-model term ${\cal L}^{({2})}$, see Eq.\ (14),
where the latter
does not contribute at all to the central interaction.
If we accept the line of reasoning above, the central interactions
from the flavor and chirally symmetric terms of the lagrangian are
viable to modifications, the others not.

In Figs.\  3-4 we show the central interactions from the flavor
symmetric
part of the lagrangian for the $N\!N$-, $\Lambda \Lambda$-, $\Sigma \Sigma$-
and $\Xi \Xi$-systems.
The general absence of any attraction is as conspicuous as the fact, that
the variation with respect to decreasing hypercharge of the system is small.
This suggests, that any improvement on the central attraction in the
$N\!N$-system will lead to very similar attractions for the other baryons.

We explore this hypothesis in Figs.\ 3-4 by varying the higher order
stabilizing terms of the lagrangian: Fig.\ 3 shows the changes when
the usual Skyrme term ${\cal L}^{({\rm 4A})}$, i.e. $\chi=0$ in
Eq.\ (\ref{lsym}), is replaced by its alternate form ${\cal
L}^{({\widetilde{\rm 4A}})}$, i.e. $\chi=1$ in Eq.\ (\ref{lsym}). This
substitution does not lead to any changes within SU(2),
but in SU(3) the situation
changes drastically as may be seen from the figure.
Since the $B=1$ solitons in the SU(2)-subgroup are
unaffected by variations of $\chi$ it is clear that
large negative values of $\chi$ will
yield arbitrary attraction for $B=2$-configurations in SU(3). But it
seems very likely to us, that the $B=1$ soliton would not be stable
towards deformations out of the SU(2)-subgroup then, to which it was
constrained by ansatz. We therefore refrain from such manipulations.
In any case, from the mesonic sector, ${\cal L}^{({\widetilde{\rm
4A}})}$ seems to be small\cite{GL85} since it violates the
Zweig rule.

Fig.\  4 shows the changes, when the fourth order stabilizer
is replaced by a sixth order one and when scalar degrees of freedom are
taken into account. For the latter, we choose a form inspired by, but
simpler than, a local approximation to the
dilatons\cite{Sch86,GJJS86,YSH89}, namely
\begin{eqnarray}
{\cal L}^{(2)} (U) & \rightarrow &
\widetilde {\cal L}^{(2)} (U) \nonumber\\
& = &
- \frac{f^2_\pi m_\pi^2}{4\kappa^2} \;
(1-\exp\bigl({-\frac{\kappa^2}{m_\pi^2}{\rm tr}\; L_{\mu}
L^{\mu}})\bigr),
\end{eqnarray}
which effectively scales down $f_\pi$ in the interior of the soliton.
Expanding the exponential to second order gives the non-linear
$\sigma$-model and the usual attractive
symmetric fourth order term which destabilizes the soliton, but here
there are higher order terms present in the exponential to counteract
this destabilization now.
Within SU(2) the inclusion of dilatons was shown to lead to an
intermediate range central attraction\cite{YSH89} when these fields were
made strong enough. In SU(3), however, strong dilaton fields lead to
a problem since not only $f_\pi$ is scaled down in the interior but
also all mass terms resulting in too small hyperon splittings in the
$B=1$ sector\cite{KSS90}.

As already mentioned, the rationale of all these different higher
order terms is not only to investigate further mechanisms for an
intermediate range attraction, but also to
demonstrate that the central interaction from the flavor symmetric
terms is fairly independent of the hypercharge of the two-baryon
configuration. Thus, if the $N\!N$-system acquires it's intermediate
range attraction from the flavor symmetric terms, this attraction will also be
present for the hyperons. It remains to be said, that the
interactions in Figs.\  3-4 have all been determined with baryon
wavefunctions diagonalizing the rotational motion in the presence of
SU(3) symmetry breaking, see Eq.\ (A1).
The distortions in the baryon wavefunctions
away from SU(3), however, do not play a decisive role here, which
means, that SU(3) symmetric wavefunctions or wavefunctions in the
strong symmetry breaking limit always lead to very small variations
of the flavor symmetric central interaction with respect to hypercharge.

Fig.\  5 shows the central interactions for identical baryons as they
arise from those terms, which explicitly break the flavor symmetry.
These contributions arise solely from the Euler angle matrix elements
of
\begin{eqnarray*}
U_0^{({\rm FSB})} (R)
\Bigl(1 - D_{88} (A)\Bigr)
\Bigl(1 - D_{88} (B)\Bigr)
\end{eqnarray*}
in Eq.\ (12) and are directly proportional to the product of the
strangeness contents of the two interacting baryons, since
$\langle Y \vert \bar s s \vert Y \rangle =
\langle Y \vert \case{1}{3}\Bigl(1 - D_{88} \Bigr) \vert Y \rangle$.
{}From Fig.\  5 we see that $U_0^{({\rm FSB})} (R)$ is attractive and
large, such that binding is enhanced with growing strangeness content.
Since this additional binding comes from a term fixed in the
mesonic sector, we judge this to be a reliable conclusion, which also
summarises one of the main points of the present work.

In Appendix  \ref{secappB}  we give a complete list of
all interaction components for
$N\!N$, $N\!\Lambda$, $N\Sigma$, $N\Xi$, $\Lambda\Lambda$, $\Sigma\Sigma$
and
$\Xi\Xi$ systems using only the normal Skyrme term, ${\cal L}^{({\rm 4A})}$,
as stabilizer adjusting the $N\Delta$-splitting.
Again, the longer-range spin-isospin dependence is
due to terms fixed in the mesonic sector and thus insensitive to any
reasonable modifications of the lagrangian. For the $N\!N$-case we
exploit one of the advantages of our approach to the
SU(3)-interactions, namely that we are able to switch continuously
from SU(3)-symmetry to the SU(2)-limit by simply changing the
baryon wavefunctions, see Eq.\ (A1).
Comparing these cases
for the $N\!N$-interaction, we find that there is a
smooth transition
with small changes in the longer range parts of the interaction to the
well known results in SU(2). This
finding  apparently is in contrast to other work published recently
on the baryon-baryon
interactions in the SU(3) Skyrme model\cite{KE90,KE92}  and
 similar discrepancies
persist for other baryon pairs, too,  which could be\cite{Kal94}
due to difficulties in the Monte Carlo integrations used in
ref.\cite{KE90,KE92}.

{}From the fact that two nucleons are loosely bound and well separated
in an isospin $I=0$, spin
$J=1$ state (the deuteron), and that no additional repulsion arises in
the
longer-range central interactions when hypercharge is decreased, we
are able to predict the spin and isospin of bound deuteron-like states
of hyperons. To this end we just have to examine the sign of the
different spin-isospin components of the long-range interaction,
which, as already has been stated, is due to well established terms
in the lagrangian. Explicitly (see Appendix  \ref{secappB}) we find that an
$N\!\Lambda$-pair should be bound in a ($I=\case{1}{2},\; J=1$)-state,
an $N\Sigma$-pair in a ($I=\case{1}{2},\; J=1$)-state,
an $N\Xi$-pair in a ($I=0,\; J=0$)-state,
a $\Lambda\Lambda$-pair in a ($I=0,\; J=0$)-state,
a $\Sigma\Sigma$-pair in
a ($I=2,\; J=0$)-state,
and a $\Xi\Xi$-pair in a ($I=1,\; J=0$)-state.

In SU(2) the most attractive configuration, the torus, can be found
numerically by initializing a general minimization algorithm
with the most attractive
product configuration\cite{KS87,VWWW86,WW92}. This most attractive product
configuration occurs at a relative orientation of
$C=\exp(i\frac{\pi}{2}\lambda_2)$ for
the two hedgehogs. With the tools set up for the present
investigation we are in a position to determine the most attractive
product configuration in SU(3) by minimizing the flavor symmetric
part of the interaction energy, Eq.\ (7) with respect to the 8 Euler
angles in $C$ at all separations.
Interestingly, the most attractive orientation is unchanged, i.e. it
resides in the SU(2) subgroup, hinting that the torus will also be the
minimum
in the $B=2$-sector of SU(3). In case of flavor symmetry with all
meson masses equal to the pion mass the torus is known\cite{SSG93}
to be lower in energy than
the SO(3)-soliton found by Balachandran {\it et al.}\cite{BLRS85}.

\section{Conclusions}
\label{sec5}

In the present work we have explored to which extent it
is possible to make reliable statements about whether hyperons
form bound deuteron like states.  To this end we have used the SU(3)
Skyrme model and compared the $N\!N$-interaction in the product ansatz
to different $NY$- and $YY$-interactions which emerge
from the same ansatz.  ($Y$ represents any octet baryon with
non-zero strangeness.)  For loosely bound deuteron like configurations
the product ansatz is sufficient if the only purpose
is the investigation of whether there is more binding in the
$NY$-system relative to the $N\!N$-case in the longer range part of
the interaction.

Such a comparison within the Skyrme model unfortunately also meets two
other obstructions: the well-known problem of the missing intermediate
range central attraction and the incomplete knowledge on the correct
combination of higher order stabilizing terms.  The latter statement,
of course, reflects the belief that the lower order
terms are fixed by their empirical values in the mesonic sector for
which evidence has been given in several
places\cite{WSP90,Wal91,S92,M93}. Fortunately, it
turns out that the higher order stabilizers, which exclusively make up
the chiral and flavor symmetric contributions to the central
interaction, show little variation with respect to different baryon
pairs considered.  Therefore, the uncertainties associated with these
terms are eliminated when we look at differences
between $N\!N$- $NY$- and $YY$-interactions: if some combination of
higher order terms -- most likely a coupling to scalar degrees of
freedom -- leads to a central attraction in the $N\!N$-case, it will
also do so for the other cases.

There remain the longer range contributions from the
known lower order terms: the non-linear $\sigma$-model is responsible
for the spin and flavor dependent components of the interaction,
whereas the chiral and flavor symmetry breaking terms introduce
further central interactions which become increasingly attractive as
the strangeness contents of the participating baryons grow.  The
conclusion, that hyperons must be bound in deuteron like states,
therefore seems inescapable.  Although our findings
confirm or extend what is commonly expected, we think
that we actually have added some weight to such expectations.

Since the contribution of the non-linear $\sigma$-model to the
potential is also considered a reliable quantity here, we were
actually in a position to predict the spins and isospins of the most
attractive baryon-baryon configurations.  Again, the Skyrme model
result seems to agree with conclusions drawn from one boson exchange
models.

\section{Acknowledgements}
\label{sec6}

We thank A. Hayashi for allowing us to upgrade his SU(2)
$N\!N$-interaction program for the present investigation and for useful
discussions. One of us, B.S., wishes to thank the ITP at the University
of Stellenbosch for its hospitality and a very enjoyable stay.

\widetext

\newpage
\appendix
\section{}
\label{secappA}

The interaction terms depend on the Euler angles of the participating
baryons, with the angles in $A$ corresponding to the degrees of
freedom of the first baryon, the angles in $B$ to the second.  The
wavefunction for an individual spin one-half baryon can be expressed
as follows:
\begin{eqnarray}
\langle A \vert {i\;y \choose i_3}{\case{1}{2} \, 1 \choose j_3}
\rangle= \sum_{\{m\}}\sqrt{m} \;X^{iy}_{\{m\}}\;
(-)^{\case{1}{2} +j_3}
D^{\{m\}^*}_{{i\;y \choose i_3}{{\frac{1}{2}}\;1 \choose -j_3}}(A) \;
, \end{eqnarray}
where the $X_{\{m\}}^{iy}$ are real amplitudes that diagonalize the
rotational hamiltonian of the baryon considered \cite{YA88,WSP90,PSW91}.

We begin with the direct matrix elements between an incoming
baryon $X(A)$ interacting with a baryon $Y(B)$.
The relative rotations $C=A^\dagger B$ contained in the $D$-functions of
all flavor symmetric interaction terms translate into the following
expression:
\begin{eqnarray}
& &\langle X'\, Y' \vert \;
 D^{\{n\}}_{{j_1\;y_1 \choose m_1}
{j_2\;y_2 \choose m_2}}(C)\;
\vert X\, Y \rangle = \\& & \qquad \qquad\qquad \qquad
\sum_{I,y,M}
\langle X \vert \;
D^{\{n\}}_{{I\;y \choose M}{j_1\;y_1 \choose m_1}}(A)\;
\vert X' \rangle^* \,
\langle Y' \vert \;
D^{\{n\}}_{{I\;y \choose M}{j_2\;y_2 \choose m_2}}(B) \;
\vert Y \rangle \; \nonumber.
\end{eqnarray}
With the conventions of de Swart\cite{DeSwart} individual angular
integrals
are actually real, so we may omit the complex conjugation later.
For the direct matrix elements the transferred hypercharge $y$ vanishes,
and $I,j_1,j_2$ are the possible isospins and spins in the
irreducible representation $\{n\}$ corresponding to hypercharge
$y,y_1,y_2$ respectively.
The angular integrals of the baryon-baryon interaction may all be decomposed
to integrals over three $D$-functions, where isospin and spin dependence
factorizes via SU(2) Clebsch-Gordan coefficients leaving isoscalar
SU(3)-factors\cite{DeSwart}.
We define reduced SU(3) matrix elements in terms of these isoscalar
factors: \begin{eqnarray}
& &\langle (i'\;y') (\case{1}{2}\;1) \Vert D_{(IY)(J0)}^{\{n\}}
\Vert (i\;y) (
\case{1}{2}\;1) \rangle = \\
& & \qquad \qquad \quad \sum_{ \{m\},\{m'\} }
\sqrt{ \frac{m}{m'} }
X_{\{m\}}^{iy} X_{\{m'\}}^{i'y'} \sum_{\gamma}
\left(
\begin{array}{cc|c}
\{m\} & \{n\}& \{m'\}_{\gamma}\\[-2pt]
iy   &IY     & i'y'
\end{array}
\right)
\left(
\begin{array}{cc|c}
\{m\} & \{n\}& \{m'\}_{\gamma}\\[-2pt]
\case{1}{2}\;1   &  J0       & \case{1}{2}\;1
\end{array}
\right) \, ,\nonumber
\end{eqnarray}
and abbreviate the SU(2)-Clebsch-Gordan coefficients by isospin and
spin operators of rank $I=1,2$, $J=1$. For $I=1$ we adopt the usual notation
\begin{equation}
\langle b \vert T_a \vert c \rangle = (T_a)_{bc} = \left\{
\begin{array}{lll}
0&{\rm for} & \Lambda\\[1mm]
(\tau_a)_{bc} & {\rm for} & N {\rm ~and~} \Xi\\[1mm]
i \varepsilon_{abc} & {\rm for} & \Sigma
\end{array}
\right.
\end{equation}
For $I=2$ it is also convenient to  define an isospin quadrupole
operator which will appear in the $\Sigma\Sigma$-interactions
\begin{equation}
\left(
\begin{array}{cc|c}
1 & 2 & 1\\[-2pt]
c & a & b
\end{array}
\right) = (Q_a)_{bc} \, .
\end{equation}
Since all baryons considered have spin one-half, the Pauli-matrices
$\sigma_b$ suffice to reproduce all  $J=1$ matrix elements.

The different non-vanishing Euler angle matrix elements of the
interaction which
now can be expressed by the reduced matrix elements given in (A3)
and by isoscalar factors are
listed below.  The spin and isospin operators
embody the different possibilities for spin or isospin flips between
incoming and the outgoing baryons $X'(A)$ and $Y'(B)$.
Note that the implicit summations run over $i,j\in\{1,2,3\}$.
\begin{eqnarray}
& &\langle X' \, Y' \vert \Bigl(1-\lambda D_{88}(A)\Bigr)
\Bigl( 1-\lambda' D_{88}(B) \Bigr)
\vert X \, Y \rangle
\rightarrow \\& &\qquad \qquad \qquad \qquad \qquad
\Bigl(1-\lambda\langle ( i_Xy_X) (\case{1}{2} 1) \Vert
D_{(00)(00)}^{\{ 8 \}}\Vert (i_Xy_X)
(\case{1}{2}1) \rangle\Bigr) \,
\nonumber \\ & & \qquad \qquad \qquad\qquad \qquad \qquad\qquad \qquad
\Bigl( 1-\lambda'\langle (i_Yy_Y) (\case{1}{2} 1) \Vert
D_{(00)(00)}^{\{ 8 \}}\Vert (i_Yy_Y) (\case{1}{2}1) \rangle \Bigr)
\nonumber
\end{eqnarray}
for arbitrary coefficients $\lambda, \lambda'$.
\begin{eqnarray}
\langle X' \, Y' \vert D_{j8}(A) D_{j8}(B) \vert X \, Y \rangle
\rightarrow & &
   {\mbox {\boldmath $T$}}_X \cdot {\mbox {\boldmath $T$}}_Y
\frac{\sqrt{(2i_X + 1)(2i_Y +1)}}{6}\nonumber \\
& & \qquad
\langle ( i_Xy_X) (\case{1}{2} 1) \Vert
D_{(10)(00)}^{\{ 8 \}}\Vert (i_Xy_X)
(\case{1}{2}1) \rangle \, \nonumber \\
& & \qquad \qquad \qquad \langle (i_Yy_Y) (\case{1}{2} 1) \Vert
D_{(10)(00)}^{\{ 8 \}}\Vert (i_Yy_Y) (\case{1}{2}1) \rangle
\end{eqnarray}
\begin{eqnarray}
\langle X' \, Y' \vert D_{8i}(A) D_{8i}(B) \vert X \, Y \rangle
\rightarrow & & \case{1}{3}
{\mbox {\boldmath $\sigma$}}_X \cdot {\mbox {\boldmath $\sigma$}}_Y
\langle ( i_Xy_X) (\case{1}{2} 1) \Vert
D_{(00)(10)}^{\{ 8 \}}\Vert (i_Xy_X)
(\case{1}{2}1) \rangle \, \nonumber \\
& & \qquad \qquad \qquad \langle (i_Yy_Y) (\case{1}{2} 1) \Vert
D_{(00)(10)}^{\{ 8 \}}\Vert (i_Yy_Y) (\case{1}{2}1) \rangle
\end{eqnarray}
\begin{eqnarray}
\langle X' \, Y' \vert D_{ji}(A) D_{ji}(B) \vert X \, Y \rangle
\rightarrow & & \case{1}{3}
{\mbox {\boldmath $\sigma$}}_X \cdot {\mbox {\boldmath $\sigma$}}_Y
  \; {\mbox {\boldmath $T$}}_X \cdot {\mbox {\boldmath $T$}}_Y
\frac{\sqrt{(2i_X + 1)(2i_Y +1)}}{6}\nonumber \\
& & \qquad
\langle ( i_Xy_X) (\case{1}{2} 1) \Vert
D_{(10)(10)}^{\{ 8 \}}\Vert (i_Xy_X)
(\case{1}{2}1) \rangle \, \nonumber \\
& & \qquad \qquad \qquad \langle (i_Yy_Y) (\case{1}{2} 1) \Vert
D_{(10)(10)}^{\{ 8 \}}\Vert (i_Yy_Y) (\case{1}{2}1) \rangle
\end{eqnarray}
The combination of $D$-functions where the right indices are coupled to
a second rank tensor
\begin{eqnarray*}
3 D_{83}(A) D_{83}(B) - D_{8i}(A) D_{8i}(B) \\
3 D_{j3}(A) D_{j3}(B) - D_{ji}(A) D_{ji}(B)
\end{eqnarray*}
lead to the same expressions as (A8) and (A9) respectively apart from the
replacement
${\mbox {\boldmath $\sigma$}}_X \cdot {\mbox {\boldmath $\sigma$}}_Y
\rightarrow S_{XY} $ where
\begin{eqnarray}
 S_{XY}= 3{\mbox {\boldmath $\sigma$}}_X \cdot \hat R \;
{\mbox {\boldmath $\sigma$}}_Y \cdot \hat R
-{\mbox {\boldmath $\sigma$}}_X \cdot {\mbox {\boldmath $\sigma$}}_Y
\end{eqnarray}
is the tensor force operator and $\hat R ={\bf e}_3$ has been used.

\begin{eqnarray}
& &\langle X' \, Y' \vert \case{1}{9} Q^{(I)}
D_{ji}(C) D_{ji}(C) \vert X \, Y \rangle
\rightarrow
\case{1}{3} \sum_{\{ n \}{\gamma}}
\left(
\begin{array}{cc|c}
\{ 8 \} & \{ 8 \} & \{ n \}_{\gamma}\\[-2pt]
10      & 10      & 00
\end{array}
\right)^2 q_{\{n\}}^{(I)}\times
\\ & & \quad
\Bigl\{
\langle ( i_Xy_X) (\case{1}{2} 1) \Vert
D_{(00)(00)}^{\{ n \}}\Vert (i_Xy_X)
(\case{1}{2}1) \rangle
\langle (i_Yy_Y) (\case{1}{2} 1) \Vert
D_{(00)(00)}^{\{ n \}}\Vert (i_Yy_Y) (\case{1}{2}1) \rangle  \nonumber
\\
& & \qquad + {\mbox {\boldmath $T$}}_X \cdot {\mbox {\boldmath $T$}}_Y
\frac{\sqrt{(2i_X + 1)(2i_Y +1)}}{6}
\langle ( i_Xy_X) (\case{1}{2} 1) \Vert
D_{(10)(00)}^{\{ n \}}\Vert (i_Xy_X)
(\case{1}{2}1) \rangle  \nonumber \\
& & \qquad \qquad \qquad \qquad \qquad \qquad\qquad \qquad \qquad
\langle (i_Yy_Y) (\case{1}{2}1) \Vert
D_{(10)(00)}^{\{ n \}}\Vert (i_Yy_Y) (\case{1}{2}1) \rangle  \nonumber
\\
& & \qquad +{\mbox {\boldmath $Q$}}_X \cdot {\mbox {\boldmath $Q$}}_Y
\langle ( i_Xy_X) (\case{1}{2} 1) \Vert
D_{(20)(00)}^{\{ n \}}\Vert (i_Xy_X)
(\case{1}{2}1) \rangle
\langle (i_Yy_Y) (\case{1}{2}1) \Vert
D_{(20)(00)}^{\{ n \}}\Vert (i_Yy_Y) (\case{1}{2}1) \rangle
\Bigr\} \nonumber
\end{eqnarray}
\begin{eqnarray}
& &\langle X' \, Y' \vert \case{1}{4} Q^{(I)} \Bigl(D_{ii}(C) D_{jj}(C)
- D_{ij}(C) D_{ji}(C) \Bigr) \vert X \, Y \rangle
\rightarrow  \\
& &\quad \case{1}{3}
{\mbox {\boldmath $\sigma$}}_X \cdot {\mbox {\boldmath $\sigma$}}_Y
\case{1}{2}
\sum_{\{ n \}{\gamma}}
\left(
\begin{array}{cc|c}
\{ 8 \} & \{ 8 \} & \{ n \}_{\gamma}\\[-2pt]
10      & 10      & 10
\end{array}
\right)^2 q_{\{n\}}^{(I)}\times
\nonumber \\ & &\qquad
\Bigl\{
\langle ( i_Xy_X) (\case{1}{2} 1) \Vert
D_{(00)(10)}^{\{ n \}}\Vert (i_Xy_X)
(\case{1}{2}1) \rangle
\langle (i_Yy_Y) (\case{1}{2}1) \Vert
D_{(00)(10)}^{\{ n \}}\Vert (i_Yy_Y) (\case{1}{2}1) \rangle  \nonumber
\\
& & \qquad \qquad+ {\mbox {\boldmath $T$}}_X \cdot {\mbox {\boldmath $T$}}_Y
\frac{\sqrt{(2i_X + 1)(2i_Y +1)}}{6}
\langle ( i_Xy_X) (\case{1}{2} 1) \Vert
D_{(10)(10)}^{\{ n \}}\Vert (i_Xy_X)
(\case{1}{2}1) \rangle  \nonumber \\
& & \qquad \qquad \qquad \qquad \qquad \qquad \qquad\qquad \qquad \qquad
\langle (i_Yy_Y) (\case{1}{2}1) \Vert
D_{(10)(10)}^{\{ n \}}\Vert (i_Yy_Y) (\case{1}{2}1) \rangle  \nonumber
\\
& & \qquad\qquad +{\mbox {\boldmath $Q$}}_X \cdot {\mbox {\boldmath $Q$}}_Y
\langle ( i_Xy_X) (\case{1}{2} 1) \Vert
D_{(20)(10)}^{\{ n \}}\Vert (i_Xy_X)
(\case{1}{2}1) \rangle
\langle (i_Yy_Y) (\case{1}{2}1) \Vert
D_{(20)(00)}^{\{ n \}}\Vert (i_Yy_Y) (\case{1}{2}1) \rangle
\Bigr\} \nonumber
\end{eqnarray}
As has been explained in section 3. the evaluation of antisymmetric
fourth order terms from ${\cal L}^{(4A)}$ is greatly simplified by the use
of the operator
$Q^{(I)}$, Eq.\ (8), for which we now give the explicit eigenvalues:
\begin{eqnarray}
q^{(I)}_{\{n\}}=\left\{
\begin{array}{cc}
1 & (I) =(\widetilde{4A}),(6) \\
3-\case{1}{6}(p^2+q^2+3p+3q+pq)& (I)=(4A)
\end{array}
\right. \, ,
\end{eqnarray}
$p$ and $q$ are related to the dimension of the representation $\{n\}$ via
$n=\case{1}{2}(p+1)(q+1)(p+q+2)$.
The combination of $D$-functions where the spin indices are coupled to
a second rank tensor
\begin{eqnarray*}
\case{1}{4}\Bigl(6 D_{11}(C) D_{22}(C) - 6 D_{12}(C) D_{21}(C)
- D_{ii}(C) D_{jj}(C) + D_{ij}(C) D_{ji}(C) \Bigr)
\end{eqnarray*}
leads to the same expressions as (A12) apart from the
replacement
${\mbox {\boldmath $\sigma$}}_X \cdot {\mbox {\boldmath $\sigma$}}_Y
\rightarrow S_{XY} $.

For the sake of brevity, and because no other interactions will be
considered, we give the exchange matrix elements only for the case
where one of the interacting baryons is a nucleon $N$ the other then
is a hyperon $Y$ the hypercharge of which we also designate by $Y$.
For $C=A^\dagger B$ we make use of the relation
\begin{eqnarray}
& &\langle Y'\, N' \vert
D^{\{n\}}_{{i_1\;y_1 \choose m_1}{i_2\;y_2 \choose m_2}}(C)
\vert N\, Y \rangle = \\& & \qquad \qquad\qquad \qquad
\sum_{I,y,M}
\langle N \vert
D^{\{n\}}_{{I\;y \choose M}{i_1\;y_1 \choose m_1}}(A)
\vert Y' \rangle^* \,
\langle N' \vert
D^{\{n\}}_{{I\;y \choose M}{i_2\;y_2 \choose m_2}}(B)
\vert Y \rangle \; . \nonumber
\end{eqnarray}
Again, with the conventions of de Swart\cite{DeSwart} individual
angular integrals are real, so we may omit the complex conjugation.
Non-vanishing matrix elements must now transfer hypercharge $y=1-Y$
and this also restricts the possible isospins $I$ occurring in the sum
of (A14).

We can give a compact representation of the non-vanishing exchange
matrix elements once the different isospin cases occurring are
abbreviated by
\begin{eqnarray}
{\bf T}^I_{NY }=\left\{
\begin{array}{l}
\left[
\begin{array}{ccc}
1 & {\rm ~~~~~~~~}Y=\Lambda & \\
\case{1}{3}(1+
{\mbox {\boldmath $\tau$}}_N \cdot {\mbox {\boldmath $T$}}_\Sigma)&
{\rm ~~~for~}Y=\Sigma&{\rm ~~~for~} I=\case{1}{2}\\
0&{\rm ~~~~~~~~}Y=\Xi &
\end{array}
\right. \nonumber \\ \nonumber \\
\left[
\begin{array}{ccc}
0 & {\rm ~~~~~~~~}Y=\Lambda & \\
\case{1}{3}(1-\case{1}{2}
{\mbox {\boldmath $\tau$}}_N \cdot {\mbox {\boldmath $T$}}_\Sigma)&
{\rm ~~~for~}Y=\Sigma&{\rm ~~~for~} I=\frac{3}{2}\\
0&{\rm ~~~~~~~~}Y=\Xi &
\end{array}
\right. \nonumber \\ \nonumber \\
\left[
\begin{array}{ccc}
0 & {\rm ~~~~~~~~}Y=\Lambda & \\
0&{\rm ~~~for~}Y=\Sigma&{\rm ~~~for~} I=0\\
\case{1}{2}(1+
{\mbox {\boldmath $\tau$}}_N \cdot {\mbox {\boldmath $\tau$}}_\Xi)&
{\rm ~~~~~~~~}Y=\Xi &
\end{array}
\right. \nonumber \\ \nonumber \\
\left[
\begin{array}{ccc}
0 & {\rm ~~~~~~~~}Y=\Lambda & \\
0&{\rm ~~~for~}Y=\Sigma&{\rm ~~~for~} I=1\\
\case{1}{2}(1-\case{1}{3}
{\mbox {\boldmath $\tau$}}_N \cdot {\mbox {\boldmath $\tau$}}_\Xi)&
{\rm ~~~~~~~~}Y=\Xi &
\end{array}
\right. \nonumber
\end{array}
\right.
\end{eqnarray}
As usual, we follow our conventions for implicit sums:
$i,j\in\{1,2,3\}$ and $k\in\{4,5,6,7\}$.
\begin{eqnarray}
\langle Y'\, N' \vert D_{ii}(C)  \vert N \, Y \rangle
\rightarrow & &
\case{1}{2}(1-\case{1}{3}
   {\mbox {\boldmath $\sigma$}}_N \cdot {\mbox {\boldmath $\sigma$}}_Y)\,
{\bf T}^I_{NY} \,
\langle ( \case{1}{2} 1)  (\case{1}{2} 1) \Vert
D_{(Iy)(10)}^{\{ 8 \}}\Vert (i_Yy_Y) (\case{1}{2}1) \rangle^2
\end{eqnarray}
The combination of $D$-functions where the spin indices are coupled to
a second rank tensor
\begin{eqnarray*}
3 D_{33}(C) - D_{ii}(C)
\end{eqnarray*}
leads to the same expression as (A15) apart from the
replacement
$\case{1}{2}(1-\case{1}{3}
{\mbox {\boldmath $\sigma$}}_N \cdot {\mbox {\boldmath $\sigma$}}_Y)
\rightarrow \case{1}{3} S_{NY} $.
\begin{eqnarray}
\langle Y'\, N' \vert D_{88}(C)  \vert N \, Y \rangle
\rightarrow
\case{1}{2}(1+
   {\mbox {\boldmath $\sigma$}}_N \cdot {\mbox {\boldmath $\sigma$}}_Y)\,
{\bf T}^I_{NY} \,
\langle ( \case{1}{2} 1)  (\case{1}{2} 1) \Vert
D_{(Iy)(00)}^{\{ 8 \}}\Vert (i_Yy_Y) (\case{1}{2}1) \rangle^2
\end{eqnarray}
\begin{eqnarray}
& &\langle Y'\, N' \vert \case{1}{9} Q^{(I)}
D_{ij}(C)D_{ij}(C)  \vert N \, Y \rangle
\rightarrow  \\
& &\quad \case{1}{6}(1+
   {\mbox {\boldmath $\sigma$}}_N \cdot {\mbox {\boldmath $\sigma$}}_Y)\,
\sum_{\{ n \}{\gamma}}
\left(
\begin{array}{cc|c}
\{ 8 \} & \{ 8 \} & \{ n \}_{\gamma}\\[-2pt]
10      & 10      & 00
\end{array}
\right)^2 q_{\{n\}}^{(I)}
\sum_I \langle ( \case{1}{2} 1)  (\case{1}{2} 1) \Vert
D_{(Iy)(00)}^{\{ 8 \}}\Vert (i_Yy_Y) (\case{1}{2}1) \rangle^2 \,
{\bf T}^I_{NY} \nonumber
\end{eqnarray}
\begin{eqnarray}
& &\langle Y'\, N' \vert \case{1}{4} Q^{(I)}
(D_{ii}(C)D_{jj}(C)-D_{ij}(C)D_{ji}(C))  \vert N \, Y \rangle
\rightarrow  \\
& &\quad \case{1}{4}(1-\case{1}{3}
   {\mbox {\boldmath $\sigma$}}_N \cdot {\mbox {\boldmath $\sigma$}}_Y)\,
\sum_{\{ n \}{\gamma}}
\left(
\begin{array}{cc|c}
\{ 8 \} & \{ 8 \} & \{ n \}_{\gamma}\\[-2pt]
10      & 10      & 10
\end{array}
\right)^2 q_{\{n\}}^{(I)}
\sum_I \langle ( \case{1}{2} 1)  (\case{1}{2} 1) \Vert
D_{(Iy)(10)}^{\{ 8 \}}\Vert (i_Yy_Y) (\case{1}{2}1) \rangle^2 \,
{\bf T}^I_{NY} \nonumber
\end{eqnarray}
The combination of $D$-functions where the spin indices are coupled to
a second rank tensor
\begin{eqnarray*}
\case{1}{4}\Biggl( 6 D_{11}(C) D_{22}(C)-6 D_{12}(C) D_{21}(C)
- D_{ii}(C) D_{jj}(C)+ D_{ij}(C) D_{ji}(C)\Biggr)
\end{eqnarray*}
leads to the same expression as (A18) apart from the
replacement
$\case{1}{4}(1-\case{1}{3}
{\mbox {\boldmath $\sigma$}}_N \cdot {\mbox {\boldmath $\sigma$}}_Y)
\rightarrow \case{1}{6} S_{NY} $.

The flavor symmetry breaking terms finally introduce other new structures
which may, however, be related to existing ones:
\begin{eqnarray}
\langle Y'\, N' \vert
D_{k8}(A)D_{k8}(B) \vert N \, Y \rangle
=
\langle Y'\, N' \vert
D_{88}(C) \vert N \, Y \rangle
\end{eqnarray}
\begin{eqnarray}
\langle Y'\, N' \vert
D_{ki}(A)D_{ki}(B) \vert N \, Y \rangle
=
\langle Y'\, N' \vert
D_{ii}(C) \vert N \, Y \rangle
\end{eqnarray}
\begin{eqnarray}
\langle Y'\, N' \vert
3D_{k3}(A)D_{k3}(B)-D_{ki}(A)D_{ki}(B) \vert N \, Y \rangle
=
\langle Y'\, N' \vert
3 D_{33}(C)-D_{ii}(C) \vert N \, Y \rangle \; .
\end{eqnarray}

\section{}
\label{secappB}

\renewcommand{\arraystretch}{.65}
In this appendix we list all interaction components of octet baryon
interactions for the lagrangian given by Eqs.\ (1-5,20 ). Only the
cumbersome but numerically small contributions from the term
proportional to $f_K^2-f_\pi^2$ have been dropped. The parameters are
those fitting the $B=0$ sector,
$f_\pi=93.0{\rm MeV}$, $m_\pi=138.0{\rm MeV}$, $f_K=113.5{\rm MeV}$,
$m_K=495.0{\rm MeV}$,
with the higher order terms,
$e= 4.12$, $\chi= 0$, $\kappa=0$, $\epsilon_6=0$, adjusted to the
$N\Delta$-split in the $B=1$-sector.
The Euler angle wave functions are obtained by diagonalizing the
rotational and flavor symmetry breaking terms in the rigid rotator
approximation.

The tables list the different radial functions subscripted
by the commonly used mnemonic notation indicating the
two-body spin or isospin operator combination according to (A4, A5, A10)
which must be multiplied.
In order to maintain uniformity of the tables there often are
superfluous columns in the tables, e.g. the isoquadrupole-isoquadrupole
interactions $U_{QQ}$, $U_{\sigma\sigma QQ}$, $U_{SQQ}$ which can only
be non-zero for the $\Sigma \Sigma$-system. For the
$\Sigma \Sigma$-system they then turn out to be numerically small.

\mediumtext
\begin{table}
\caption{}
\begin{tabular}{cdccdccccc}
 \multicolumn{10}{c}{ $N\!N$ INTERACTION}\\
 \multicolumn{1}{c}{R$\!$ [fm]}
& \multicolumn{1}{c}{$U_0$}
& \multicolumn{1}{c}{$U_{\sigma\sigma}$}
& \multicolumn{1}{c}{$U_{\tau\tau}$}
& \multicolumn{1}{c}{$U_{\sigma\sigma\tau\tau}$}
& \multicolumn{1}{c}{$U_S$}
& \multicolumn{1}{c}{$U_{S\tau\tau}$}
& \multicolumn{1}{c}{$U_{QQ}$}
& \multicolumn{1}{c}{$U_{\sigma\sigma QQ}$}
& \multicolumn{1}{c}{$U_{SQQ}$} \\
\hline
  0.10&  722.1&    0.3&    0.3&   15.9&    0.0&    0.2&    0.0&    0.0&
0.0\\
  0.20&  680.9&    0.3&    0.3&   15.2&    0.0&    0.6&    0.0&    0.0&
0.0\\
  0.30&  618.1&    0.3&    0.2&   14.1&    0.0&    1.2&    0.0&    0.0&
0.0\\
  0.40&  540.9&    0.3&    0.2&   12.8&    0.0&    1.8&    0.0&    0.0&
0.0\\
  0.50&  457.0&    0.2&    0.2&   11.3&    0.0&    2.5&    0.0&    0.0&
0.0\\
  0.60&  373.2&    0.2&    0.1&    9.8&    0.0&    3.0&    0.0&    0.0&
0.0\\
  0.70&  295.1&    0.2&    0.1&    8.3&    0.0&    3.3&    0.0&    0.0&
0.0\\
  0.80&  226.3&    0.1&    0.1&    6.9&    0.0&    3.5&    0.0&    0.0&
0.0\\
  0.90&  168.5&    0.1&    0.1&    5.5&    0.0&    3.5&    0.0&    0.0&
0.0\\
  1.00&  122.2&    0.1&    0.0&    4.4&    0.0&    3.4&    0.0&    0.0&
0.0\\
  1.10&   86.6&    0.1&    0.0&    3.4&    0.0&    3.1&    0.0&    0.0&
0.0\\
  1.20&   60.1&    0.1&    0.0&    2.6&    0.0&    2.8&    0.0&    0.0&
0.0\\
  1.30&   41.1&    0.1&    0.0&    2.0&    0.0&    2.5&    0.0&    0.0&
0.0\\
  1.40&   27.8&    0.0&    0.0&    1.5&    0.0&    2.2&    0.0&    0.0&
0.0\\
  1.50&   18.6&    0.0&    0.0&    1.1&    0.0&    1.9&    0.0&    0.0&
0.0\\
  1.60&   12.5&    0.0&    0.0&    0.9&    0.0&    1.7&    0.0&    0.0&
0.0\\
  1.70&    8.3&    0.0&    0.0&    0.7&    0.0&    1.4&    0.0&    0.0&
0.0\\
  1.80&    5.6&    0.0&    0.0&    0.5&    0.0&    1.2&    0.0&    0.0&
0.0\\
  1.90&    3.8&    0.0&    0.0&    0.4&    0.0&    1.1&    0.0&    0.0&
0.0\\
  2.00&    2.6&    0.0&    0.0&    0.3&    0.0&    0.9&    0.0&    0.0&
0.0\\
  2.10&    1.7&    0.0&    0.0&    0.3&    0.0&    0.8&    0.0&    0.0&
0.0\\
  2.20&    1.2&    0.0&    0.0&    0.2&    0.0&    0.7&    0.0&    0.0&
0.0\\
  2.30&    0.8&    0.0&    0.0&    0.2&    0.0&    0.6&    0.0&    0.0&
0.0\\
  2.40&    0.6&    0.0&    0.0&    0.2&    0.0&    0.5&    0.0&    0.0&
0.0\\
  2.50&    0.4&    0.0&    0.0&    0.1&    0.0&    0.4&    0.0&    0.0&
0.0\\
  3.00&    0.1&    0.0&    0.0&    0.1&    0.0&    0.2&    0.0&    0.0&
0.0\\
  3.50&    0.0&    0.0&    0.0&    0.0&    0.0&    0.1&    0.0&    0.0&
0.0\\
\end{tabular}
\label{table1}
\end{table}

\begin{table}
\caption{}
\begin{tabular}{cdccdccccc}
 \multicolumn{10}{c}{ $N\!\Lambda$ INTERACTION}\\
 \multicolumn{1}{c}{R$\!$ [fm]}
& \multicolumn{1}{c}{$U_0$}
& \multicolumn{1}{c}{$U_{\sigma\sigma}$}
& \multicolumn{1}{c}{$U_{\tau\tau}$}
& \multicolumn{1}{c}{$U_{\sigma\sigma\tau\tau}$}
& \multicolumn{1}{c}{$U_S$}
& \multicolumn{1}{c}{$U_{S\tau\tau}$}
& \multicolumn{1}{c}{$U_{QQ}$}
& \multicolumn{1}{c}{$U_{\sigma\sigma QQ}$}
& \multicolumn{1}{c}{$U_{SQQ}$} \\
\hline
  0.10&  689.8&   -7.5&    0.0&    0.0&    0.0&    0.0&    0.0&    0.0&
0.0\\
  0.20&  650.8&   -7.4&    0.0&    0.0&   -0.1&    0.0&    0.0&    0.0&
0.0\\
  0.30&  591.2&   -7.3&    0.0&    0.0&   -0.3&    0.0&    0.0&    0.0&
0.0\\
  0.40&  518.0&   -7.1&    0.0&    0.0&   -0.6&    0.0&    0.0&    0.0&
0.0\\
  0.50&  438.4&   -6.8&    0.0&    0.0&   -0.9&    0.0&    0.0&    0.0&
0.0\\
  0.60&  358.9&   -6.4&    0.0&    0.0&   -1.2&    0.0&    0.0&    0.0&
0.0\\
  0.70&  284.7&   -5.9&    0.0&    0.0&   -1.6&    0.0&    0.0&    0.0&
0.0\\
  0.80&  219.1&   -5.4&    0.0&    0.0&   -1.9&    0.0&    0.0&    0.0&
0.0\\
  0.90&  164.0&   -4.9&    0.0&    0.0&   -2.2&    0.0&    0.0&    0.0&
0.0\\
  1.00&  119.6&   -4.3&    0.0&    0.0&   -2.5&    0.0&    0.0&    0.0&
0.0\\
  1.10&   85.4&   -3.8&    0.0&    0.0&   -2.7&    0.0&    0.0&    0.0&
0.0\\
  1.20&   59.8&   -3.2&    0.0&    0.0&   -2.8&    0.0&    0.0&    0.0&
0.0\\
  1.30&   41.4&   -2.7&    0.0&    0.0&   -2.9&    0.0&    0.0&    0.0&
0.0\\
  1.40&   28.4&   -2.3&    0.0&    0.0&   -2.9&    0.0&    0.0&    0.0&
0.0\\
  1.50&   19.4&   -1.9&    0.0&    0.0&   -2.9&    0.0&    0.0&    0.0&
0.0\\
  1.60&   13.3&   -1.6&    0.0&    0.0&   -2.9&    0.0&    0.0&    0.0&
0.0\\
  1.70&    9.1&   -1.3&    0.0&    0.0&   -2.8&    0.0&    0.0&    0.0&
0.0\\
  1.80&    6.3&   -1.1&    0.0&    0.0&   -2.7&    0.0&    0.0&    0.0&
0.0\\
  1.90&    4.4&   -0.9&    0.0&    0.0&   -2.6&    0.0&    0.0&    0.0&
0.0\\
  2.00&    3.1&   -0.7&    0.0&    0.0&   -2.4&    0.0&    0.0&    0.0&
0.0\\
  2.10&    2.2&   -0.6&    0.0&    0.0&   -2.3&    0.0&    0.0&    0.0&
0.0\\
  2.20&    1.6&   -0.5&    0.0&    0.0&   -2.2&    0.0&    0.0&    0.0&
0.0\\
  2.30&    1.2&   -0.4&    0.0&    0.0&   -2.1&    0.0&    0.0&    0.0&
0.0\\
  2.40&    0.9&   -0.3&    0.0&    0.0&   -1.9&    0.0&    0.0&    0.0&
0.0\\
  2.50&    0.6&   -0.2&    0.0&    0.0&   -1.8&    0.0&    0.0&    0.0&
0.0\\
  3.00&    0.1&    0.0&    0.0&    0.0&   -1.3&    0.0&    0.0&    0.0&
0.0\\
  3.50&    0.0&    0.0&    0.0&    0.0&   -0.9&    0.0&    0.0&    0.0&
0.0\\
\end{tabular}
\label{table2}
\end{table}

\begin{table}
\caption{}
\begin{tabular}{cdccdccccc}
 \multicolumn{10}{c}{ $N\Sigma$ INTERACTION}\\
 \multicolumn{1}{c}{R$\!$ [fm]}
& \multicolumn{1}{c}{$U_0$}
& \multicolumn{1}{c}{$U_{\sigma\sigma}$}
& \multicolumn{1}{c}{$U_{\tau\tau}$}
& \multicolumn{1}{c}{$U_{\sigma\sigma\tau\tau}$}
& \multicolumn{1}{c}{$U_S$}
& \multicolumn{1}{c}{$U_{S\tau\tau}$}
& \multicolumn{1}{c}{$U_{QQ}$}
& \multicolumn{1}{c}{$U_{\sigma\sigma QQ}$}
& \multicolumn{1}{c}{$U_{SQQ}$} \\
\hline
  0.10&  628.7&   12.3&   94.5&  100.5&    0.0&    0.1&    0.0&    0.0&
0.0\\
  0.20&  592.6&   11.5&   89.2&   94.9&    0.0&    0.5&    0.0&    0.0&
0.0\\
  0.30&  537.6&   10.4&   81.1&   86.4&    0.0&    1.0&    0.0&    0.0&
0.0\\
  0.40&  470.0&    9.1&   71.1&   76.0&   -0.1&    1.5&    0.0&    0.0&
0.0\\
  0.50&  396.6&    7.6&   60.3&   64.6&   -0.1&    2.0&    0.0&    0.0&
0.0\\
  0.60&  323.5&    6.2&   49.4&   53.3&   -0.1&    2.4&    0.0&    0.0&
0.0\\
  0.70&  255.4&    4.8&   39.3&   42.6&   -0.2&    2.7&    0.0&    0.0&
0.0\\
  0.80&  195.4&    3.7&   30.3&   33.1&   -0.2&    2.8&    0.0&    0.0&
0.0\\
  0.90&  145.1&    2.7&   22.8&   25.0&   -0.3&    2.8&    0.0&    0.0&
0.0\\
  1.00&  104.8&    1.9&   16.7&   18.4&   -0.3&    2.7&    0.0&    0.0&
0.0\\
  1.10&   73.9&    1.3&   12.0&   13.3&   -0.3&    2.5&    0.0&    0.0&
0.0\\
  1.20&   51.0&    0.9&    8.4&    9.5&   -0.4&    2.2&    0.0&    0.0&
0.0\\
  1.30&   34.6&    0.6&    5.9&    6.6&   -0.4&    2.0&    0.0&    0.0&
0.0\\
  1.40&   23.2&    0.4&    4.1&    4.6&   -0.4&    1.7&    0.0&    0.0&
0.0\\
  1.50&   15.4&    0.2&    2.8&    3.2&   -0.4&    1.5&    0.0&    0.0&
0.0\\
  1.60&   10.2&    0.1&    1.9&    2.2&   -0.4&    1.2&    0.0&    0.0&
0.0\\
  1.70&    6.7&    0.1&    1.3&    1.6&   -0.4&    1.1&    0.0&    0.0&
0.0\\
  1.80&    4.4&    0.0&    0.9&    1.1&   -0.4&    0.9&    0.0&    0.0&
0.0\\
  1.90&    2.9&    0.0&    0.7&    0.8&   -0.4&    0.7&    0.0&    0.0&
0.0\\
  2.00&    2.0&    0.0&    0.5&    0.6&   -0.3&    0.6&    0.0&    0.0&
0.0\\
  2.10&    1.3&    0.0&    0.3&    0.4&   -0.3&    0.5&    0.0&    0.0&
0.0\\
  2.20&    0.9&    0.0&    0.3&    0.3&   -0.3&    0.4&    0.0&    0.0&
0.0\\
  2.30&    0.6&    0.0&    0.2&    0.3&   -0.3&    0.4&    0.0&    0.0&
0.0\\
  2.40&    0.4&    0.0&    0.1&    0.2&   -0.3&    0.3&    0.0&    0.0&
0.0\\
  2.50&    0.3&    0.0&    0.1&    0.2&   -0.3&    0.3&    0.0&    0.0&
0.0\\
  3.00&    0.0&    0.0&    0.0&    0.1&   -0.2&    0.1&    0.0&    0.0&
0.0\\
  3.50&    0.0&    0.0&    0.0&    0.0&   -0.1&    0.1&    0.0&    0.0&
0.0\\
\end{tabular}
\label{table3}
\end{table}

\begin{table}
\caption{}
\begin{tabular}{cdccdccccc}
 \multicolumn{10}{c}{ $N\Xi$ INTERACTION}\\
 \multicolumn{1}{c}{R$\!$ [fm]}
& \multicolumn{1}{c}{$U_0$}
& \multicolumn{1}{c}{$U_{\sigma\sigma}$}
& \multicolumn{1}{c}{$U_{\tau\tau}$}
& \multicolumn{1}{c}{$U_{\sigma\sigma\tau\tau}$}
& \multicolumn{1}{c}{$U_S$}
& \multicolumn{1}{c}{$U_{S\tau\tau}$}
& \multicolumn{1}{c}{$U_{QQ}$}
& \multicolumn{1}{c}{$U_{\sigma\sigma QQ}$}
& \multicolumn{1}{c}{$U_{SQQ}$} \\
\hline
  0.10&  552.2&   -2.9&    2.7&   -4.1&    0.0&    0.0&    0.0&    0.0&
0.0\\
  0.20&  520.3&   -2.8&    2.5&   -4.0&    0.0&   -0.2&    0.0&    0.0&
0.0\\
  0.30&  471.7&   -2.6&    2.3&   -3.7&    0.0&   -0.3&    0.0&    0.0&
0.0\\
  0.40&  412.0&   -2.4&    2.0&   -3.3&    0.0&   -0.5&    0.0&    0.0&
0.0\\
  0.50&  347.3&   -2.1&    1.7&   -3.0&    0.0&   -0.7&    0.0&    0.0&
0.0\\
  0.60&  282.8&   -1.8&    1.4&   -2.6&    0.0&   -0.8&    0.0&    0.0&
0.0\\
  0.70&  222.8&   -1.6&    1.1&   -2.2&    0.0&   -0.9&    0.0&    0.0&
0.0\\
  0.80&  170.0&   -1.3&    0.9&   -1.8&    0.0&   -0.9&    0.0&    0.0&
0.0\\
  0.90&  125.8&   -1.1&    0.6&   -1.5&    0.1&   -0.9&    0.0&    0.0&
0.0\\
  1.00&   90.5&   -0.9&    0.5&   -1.2&    0.1&   -0.9&    0.0&    0.0&
0.0\\
  1.10&   63.5&   -0.7&    0.3&   -0.9&    0.1&   -0.8&    0.0&    0.0&
0.0\\
  1.20&   43.5&   -0.6&    0.2&   -0.7&    0.1&   -0.8&    0.0&    0.0&
0.0\\
  1.30&   29.3&   -0.5&    0.2&   -0.5&    0.1&   -0.7&    0.0&    0.0&
0.0\\
  1.40&   19.4&   -0.4&    0.1&   -0.4&    0.1&   -0.6&    0.0&    0.0&
0.0\\
  1.50&   12.7&   -0.3&    0.1&   -0.3&    0.2&   -0.5&    0.0&    0.0&
0.0\\
  1.60&    8.3&   -0.3&    0.1&   -0.2&    0.2&   -0.4&    0.0&    0.0&
0.0\\
  1.70&    5.3&   -0.2&    0.0&   -0.2&    0.2&   -0.4&    0.0&    0.0&
0.0\\
  1.80&    3.4&   -0.2&    0.0&   -0.1&    0.2&   -0.3&    0.0&    0.0&
0.0\\
  1.90&    2.2&   -0.1&    0.0&   -0.1&    0.2&   -0.3&    0.0&    0.0&
0.0\\
  2.00&    1.4&   -0.1&    0.0&   -0.1&    0.2&   -0.2&    0.0&    0.0&
0.0\\
  2.10&    0.9&   -0.1&    0.0&   -0.1&    0.1&   -0.2&    0.0&    0.0&
0.0\\
  2.20&    0.6&   -0.1&    0.0&   -0.1&    0.1&   -0.2&    0.0&    0.0&
0.0\\
  2.30&    0.4&   -0.1&    0.0&   -0.1&    0.1&   -0.2&    0.0&    0.0&
0.0\\
  2.40&    0.2&    0.0&    0.0&    0.0&    0.1&   -0.1&    0.0&    0.0&
0.0\\
  2.50&    0.1&    0.0&    0.0&    0.0&    0.1&   -0.1&    0.0&    0.0&
0.0\\
  3.00&    0.0&    0.0&    0.0&    0.0&    0.1&   -0.1&    0.0&    0.0&
0.0\\
  3.50&    0.0&    0.0&    0.0&    0.0&    0.1&    0.0&    0.0&    0.0&
0.0\\
\end{tabular}
\label{table4}
\end{table}

\begin{table}
\caption{}
\begin{tabular}{cdccdccccc}
 \multicolumn{10}{c}{ $\Lambda\Lambda$ INTERACTION}\\
 \multicolumn{1}{c}{R$\!$ [fm]}
& \multicolumn{1}{c}{$U_0$}
& \multicolumn{1}{c}{$U_{\sigma\sigma}$}
& \multicolumn{1}{c}{$U_{\tau\tau}$}
& \multicolumn{1}{c}{$U_{\sigma\sigma\tau\tau}$}
& \multicolumn{1}{c}{$U_S$}
& \multicolumn{1}{c}{$U_{S\tau\tau}$}
& \multicolumn{1}{c}{$U_{QQ}$}
& \multicolumn{1}{c}{$U_{\sigma\sigma QQ}$}
& \multicolumn{1}{c}{$U_{SQQ}$} \\
\hline
  0.10&  600.8&   12.6&    0.0&    0.0&    0.0&    0.0&    0.0&    0.0&
0.0\\
  0.20&  566.2&   12.1&    0.0&    0.0&    0.0&    0.0&    0.0&    0.0&
0.0\\
  0.30&  513.4&   11.3&    0.0&    0.0&    0.1&    0.0&    0.0&    0.0&
0.0\\
  0.40&  448.5&   10.3&    0.0&    0.0&    0.1&    0.0&    0.0&    0.0&
0.0\\
  0.50&  378.1&    9.2&    0.0&    0.0&    0.1&    0.0&    0.0&    0.0&
0.0\\
  0.60&  308.0&    8.1&    0.0&    0.0&    0.0&    0.0&    0.0&    0.0&
0.0\\
  0.70&  242.8&    7.0&    0.0&    0.0&   -0.1&    0.0&    0.0&    0.0&
0.0\\
  0.80&  185.3&    5.9&    0.0&    0.0&   -0.1&    0.0&    0.0&    0.0&
0.0\\
  0.90&  137.3&    4.9&    0.0&    0.0&   -0.2&    0.0&    0.0&    0.0&
0.0\\
  1.00&   98.8&    4.0&    0.0&    0.0&   -0.4&    0.0&    0.0&    0.0&
0.0\\
  1.10&   69.4&    3.3&    0.0&    0.0&   -0.5&    0.0&    0.0&    0.0&
0.0\\
  1.20&   47.7&    2.7&    0.0&    0.0&   -0.5&    0.0&    0.0&    0.0&
0.0\\
  1.30&   32.1&    2.1&    0.0&    0.0&   -0.6&    0.0&    0.0&    0.0&
0.0\\
  1.40&   21.3&    1.7&    0.0&    0.0&   -0.7&    0.0&    0.0&    0.0&
0.0\\
  1.50&   14.0&    1.4&    0.0&    0.0&   -0.7&    0.0&    0.0&    0.0&
0.0\\
  1.60&    9.1&    1.1&    0.0&    0.0&   -0.7&    0.0&    0.0&    0.0&
0.0\\
  1.70&    5.9&    0.9&    0.0&    0.0&   -0.7&    0.0&    0.0&    0.0&
0.0\\
  1.80&    3.8&    0.7&    0.0&    0.0&   -0.7&    0.0&    0.0&    0.0&
0.0\\
  1.90&    2.5&    0.6&    0.0&    0.0&   -0.7&    0.0&    0.0&    0.0&
0.0\\
  2.00&    1.6&    0.5&    0.0&    0.0&   -0.7&    0.0&    0.0&    0.0&
0.0\\
  2.10&    1.0&    0.4&    0.0&    0.0&   -0.7&    0.0&    0.0&    0.0&
0.0\\
  2.20&    0.7&    0.3&    0.0&    0.0&   -0.6&    0.0&    0.0&    0.0&
0.0\\
  2.30&    0.4&    0.2&    0.0&    0.0&   -0.6&    0.0&    0.0&    0.0&
0.0\\
  2.40&    0.3&    0.2&    0.0&    0.0&   -0.6&    0.0&    0.0&    0.0&
0.0\\
  2.50&    0.2&    0.2&    0.0&    0.0&   -0.5&    0.0&    0.0&    0.0&
0.0\\
  3.00&    0.0&    0.0&    0.0&    0.0&   -0.4&    0.0&    0.0&    0.0&
0.0\\
  3.50&    0.0&    0.0&    0.0&    0.0&   -0.3&    0.0&    0.0&    0.0&
0.0\\
\end{tabular}
\label{table5}
\end{table}

\begin{table}
\caption{}
\begin{tabular}{cdccdccccc}
 \multicolumn{10}{c}{ $\Sigma\Sigma$ INTERACTION}\\
 \multicolumn{1}{c}{R$\!$ [fm]}
& \multicolumn{1}{c}{$U_0$}
& \multicolumn{1}{c}{$U_{\sigma\sigma}$}
& \multicolumn{1}{c}{$U_{\tau\tau}$}
& \multicolumn{1}{c}{$U_{\sigma\sigma\tau\tau}$}
& \multicolumn{1}{c}{$U_S$}
& \multicolumn{1}{c}{$U_{S\tau\tau}$}
& \multicolumn{1}{c}{$U_{QQ}$}
& \multicolumn{1}{c}{$U_{\sigma\sigma QQ}$}
& \multicolumn{1}{c}{$U_{SQQ}$} \\
\hline
  0.10&  555.7&    7.0&   23.4&   10.0&    0.0&    0.1&    0.0&    0.0&
0.0\\
  0.20&  523.4&    6.7&   22.1&    9.5&    0.0&    0.4&    0.0&    0.0&
0.0\\
  0.30&  474.2&    6.3&   20.1&    8.8&    0.0&    0.7&    0.0&    0.0&
0.0\\
  0.40&  413.8&    5.8&   17.6&    8.0&    0.0&    1.2&    0.0&    0.0&
0.0\\
  0.50&  348.3&    5.2&   14.9&    7.1&    0.0&    1.6&    0.0&    0.0&
0.0\\
  0.60&  283.2&    4.5&   12.2&    6.1&    0.0&    1.9&    0.0&    0.0&
0.0\\
  0.70&  222.6&    3.9&    9.7&    5.2&    0.0&    2.1&    0.0&    0.0&
0.0\\
  0.80&  169.4&    3.3&    7.4&    4.3&   -0.1&    2.2&    0.0&    0.0&
0.0\\
  0.90&  124.9&    2.7&    5.6&    3.5&   -0.1&    2.2&    0.0&    0.0&
0.0\\
  1.00&   89.5&    2.3&    4.0&    2.8&   -0.2&    2.1&    0.0&    0.0&
0.0\\
  1.10&   62.4&    1.8&    2.9&    2.1&   -0.3&    2.0&    0.0&    0.0&
0.0\\
  1.20&   42.4&    1.5&    2.0&    1.6&   -0.3&    1.8&    0.0&    0.0&
0.0\\
  1.30&   28.3&    1.2&    1.4&    1.2&   -0.3&    1.6&    0.0&    0.0&
0.0\\
  1.40&   18.5&    1.0&    1.0&    0.9&   -0.4&    1.4&    0.0&    0.0&
0.0\\
  1.50&   11.9&    0.8&    0.6&    0.7&   -0.4&    1.2&    0.0&    0.0&
0.0\\
  1.60&    7.6&    0.6&    0.4&    0.5&   -0.4&    1.0&    0.0&    0.0&
0.0\\
  1.70&    4.8&    0.5&    0.3&    0.4&   -0.4&    0.9&    0.0&    0.0&
0.0\\
  1.80&    3.0&    0.4&    0.2&    0.3&   -0.4&    0.8&    0.0&    0.0&
0.0\\
  1.90&    1.8&    0.3&    0.1&    0.3&   -0.4&    0.7&    0.0&    0.0&
0.0\\
  2.00&    1.1&    0.3&    0.1&    0.2&   -0.4&    0.6&    0.0&    0.0&
0.0\\
  2.10&    0.7&    0.2&    0.1&    0.2&   -0.4&    0.5&    0.0&    0.0&
0.0\\
  2.20&    0.4&    0.2&    0.0&    0.1&   -0.4&    0.4&    0.0&    0.0&
0.0\\
  2.30&    0.2&    0.1&    0.0&    0.1&   -0.3&    0.4&    0.0&    0.0&
0.0\\
  2.40&    0.1&    0.1&    0.0&    0.1&   -0.3&    0.3&    0.0&    0.0&
0.0\\
  2.50&    0.0&    0.1&    0.0&    0.1&   -0.3&    0.3&    0.0&    0.0&
0.0\\
  3.00&    0.0&    0.0&    0.0&    0.0&   -0.2&    0.1&    0.0&    0.0&
0.0\\
  3.50&    0.0&    0.0&    0.0&    0.0&   -0.2&    0.1&    0.0&    0.0&
0.0\\
\end{tabular}
\label{table6}
\end{table}

\begin{table}
\caption{}
\begin{tabular}{cdccdccccc}
 \multicolumn{10}{c}{ $\Xi\Xi$ INTERACTION}\\
 \multicolumn{1}{c}{R$\!$ [fm]}
& \multicolumn{1}{c}{$U_0$}
& \multicolumn{1}{c}{$U_{\sigma\sigma}$}
& \multicolumn{1}{c}{$U_{\tau\tau}$}
& \multicolumn{1}{c}{$U_{\sigma\sigma\tau\tau}$}
& \multicolumn{1}{c}{$U_S$}
& \multicolumn{1}{c}{$U_{S\tau\tau}$}
& \multicolumn{1}{c}{$U_{QQ}$}
& \multicolumn{1}{c}{$U_{\sigma\sigma QQ}$}
& \multicolumn{1}{c}{$U_{SQQ}$} \\
\hline
  0.10&  508.0&   25.2&   23.1&    1.1&    0.0&    0.0&    0.0&    0.0&
0.0\\
  0.20&  477.9&   24.2&   21.8&    1.1&    0.1&    0.0&    0.0&    0.0&
0.0\\
  0.30&  432.2&   22.7&   19.8&    1.0&    0.1&    0.1&    0.0&    0.0&
0.0\\
  0.40&  376.3&   20.7&   17.3&    0.9&    0.1&    0.1&    0.0&    0.0&
0.0\\
  0.50&  315.7&   18.5&   14.7&    0.8&    0.1&    0.2&    0.0&    0.0&
0.0\\
  0.60&  255.6&   16.2&   12.0&    0.7&    0.0&    0.2&    0.0&    0.0&
0.0\\
  0.70&  199.8&   14.0&    9.5&    0.6&   -0.1&    0.2&    0.0&    0.0&
0.0\\
  0.80&  151.0&   11.8&    7.3&    0.5&   -0.3&    0.3&    0.0&    0.0&
0.0\\
  0.90&  110.3&    9.9&    5.5&    0.4&   -0.5&    0.3&    0.0&    0.0&
0.0\\
  1.00&   78.1&    8.1&    4.0&    0.3&   -0.7&    0.2&    0.0&    0.0&
0.0\\
  1.10&   53.6&    6.6&    2.8&    0.2&   -0.9&    0.2&    0.0&    0.0&
0.0\\
  1.20&   35.7&    5.3&    2.0&    0.2&   -1.1&    0.2&    0.0&    0.0&
0.0\\
  1.30&   23.2&    4.3&    1.4&    0.1&   -1.2&    0.2&    0.0&    0.0&
0.0\\
  1.40&   14.6&    3.5&    0.9&    0.1&   -1.3&    0.2&    0.0&    0.0&
0.0\\
  1.50&    9.0&    2.8&    0.6&    0.1&   -1.4&    0.1&    0.0&    0.0&
0.0\\
  1.60&    5.4&    2.3&    0.4&    0.1&   -1.4&    0.1&    0.0&    0.0&
0.0\\
  1.70&    3.1&    1.8&    0.3&    0.0&   -1.5&    0.1&    0.0&    0.0&
0.0\\
  1.80&    1.7&    1.5&    0.2&    0.0&   -1.4&    0.1&    0.0&    0.0&
0.0\\
  1.90&    0.9&    1.2&    0.1&    0.0&   -1.4&    0.1&    0.0&    0.0&
0.0\\
  2.00&    0.4&    1.0&    0.1&    0.0&   -1.4&    0.1&    0.0&    0.0&
0.0\\
  2.10&    0.1&    0.8&    0.1&    0.0&   -1.3&    0.1&    0.0&    0.0&
0.0\\
  2.20&   -0.1&    0.6&    0.0&    0.0&   -1.3&    0.0&    0.0&    0.0&
0.0\\
  2.30&   -0.2&    0.5&    0.0&    0.0&   -1.2&    0.0&    0.0&    0.0&
0.0\\
  2.40&   -0.2&    0.4&    0.0&    0.0&   -1.2&    0.0&    0.0&    0.0&
0.0\\
  2.50&   -0.2&    0.3&    0.0&    0.0&   -1.1&    0.0&    0.0&    0.0&
0.0\\
  3.00&   -0.1&    0.1&    0.0&    0.0&   -0.8&    0.0&    0.0&    0.0&
0.0\\
  3.50&    0.0&    0.0&    0.0&    0.0&   -0.6&    0.0&    0.0&    0.0&
0.0\\
\end{tabular}
\label{table7}
\end{table}

\narrowtext

\newpage

\narrowtext

\begin{figure}
\caption{Mass splittings for
$\case{1}{2}^+ $ and $\frac{3}{2}^+ $ baryons.
Crosses indicate experimental splittings, dashed lines the
spectrum calculated in the slow rotator approach, dots the results from
rigidly rotated solitons. Parameters used for the rigid rotator are
$f_\pi=93.0{\rm MeV}$, $m_\pi=138.0{\rm MeV}$, $f_K=113.5{\rm MeV}$, $
m_K=495.0{\rm MeV}$,
$e= 4.12$, $\chi= 0$, $\kappa=0$, $\epsilon_6=0$ and for the slow rotator
$f_\pi=93.0{\rm MeV}$, $m_\pi=138.0{\rm MeV}$, $f_K=113.5{\rm MeV}$,
$m_K=495.0{\rm MeV}$,
$e= 3.45$, $\chi= -0.07$, $\kappa=0$, $\epsilon_6=0$.
}
\label{fig1}

\end{figure}

\begin{figure}
\caption{
Direct and exchange matrix elements for non-identical baryons.
}
\label{fig2}

\end{figure}

\begin{figure}
\caption{Central interaction from the flavor symmetric part of
the lagrangian with fourth order stabilizing terms shown for the
$N\!N$-system (thin full line),  $\Lambda
\Lambda$-system (thick full line), and $\Sigma\Sigma$-
and $\Xi\Xi$-systems (dashed line). (The latter two curves
practically coincide and only one is shown.) The
lower quadruple of curves is obtained with  the parameters
$f_\pi=93.0{\rm MeV}$, $m_\pi=138.0{\rm MeV}$, $f_K=113.5{\rm MeV}$,
$m_K=495.0{\rm MeV}$,
$e= 4.12$, $\chi= 0$, $\kappa=0$, $\epsilon_6=0$. The upper group has
${\cal L}^{({\rm 4A})}$
replaced by
${\cal L}^{({\widetilde{\rm 4A}})}$
i.e. same parameters
as before, only $\chi=1$ now.
}
\label{fig3}
\end{figure}

\begin{figure}
\caption{As Fig.\ 3, but showing the effect of alternatives to
fourth order stabilizers.
(As in Fig.\ 3  $\Sigma \Sigma$- and $\Xi \Xi$-curves
practically coincide.)  Upper quadruple of
curves with sixth order stabilizer:
$f_\pi=93.0{\rm MeV}$, $m_\pi=138.0{\rm MeV}$, $f_K=113.5{\rm MeV}$,
$m_K=495.0{\rm MeV}$,
$e= 0$, $\chi= 0$, $\kappa=0$, $\epsilon_6=.0125{\rm MeV}^{-1}$.
Lower group
shows the effect of dilatons:
$f_\pi=93.0{\rm MeV}$, $m_\pi=138.0{\rm MeV}$, $f_K=113.5{\rm MeV}$,
$m_ K=495.0{\rm MeV}$,
$e= 4.12$, $\chi= 0$, $\kappa=0.03$, $\epsilon_6=0$.
}
\label{fig4}
\end{figure}

\begin{figure}
\caption{
Central interaction from the flavor symmetry breaking part of the
lagrangian for the $N\!N$-system (thin full line), the $\Lambda
\Lambda$-system (thick full line), the $\Sigma \Sigma$-system (dotted
line) and the $\Xi \Xi$-system (dashed line). The
curves are obtained for the parameters
$f_\pi=93.0{\rm MeV}$, $m_\pi=138.0{\rm MeV}$, $f_K=113.5{\rm MeV}$,
$m_K=495.0{\rm MeV}$,
$e= 4.12$, $\chi= 0$, $\kappa=0$, $\epsilon_6=0$.
}
\label{fig5}
\end{figure}

\newpage

\widetext

\noindent Figure 1 \\
Mass splittings for
$\case{1}{2}^+ $ and $\frac{3}{2}^+ $ baryons.
The crosses indicate the experimental splittings, the dashed lines the
spectrum calculated in the slow rotator approach, dots the results from
rigidly rotated solitons. For the rigid rotator we used the parameters
$f_\pi=93.0{\rm MeV}$, $m_\pi=138.0{\rm MeV}$, $f_K=113.5{\rm MeV}$, $
m_K=495.0{\rm MeV}$,
$e= 4.12$, $\chi= 0$, $\kappa=0$, $\epsilon_6=0$ and for the slow rotator
$f_\pi=93.0{\rm MeV}$, $m_\pi=138.0{\rm MeV}$, $f_K=113.5{\rm MeV}$,
$m_K=495.0{\rm MeV}$,
$e= 3.45$, $\chi= -0.07$, $\kappa=0$, $\epsilon_6=0$.

\begin{center} \setlength{\unitlength}{1cm}
\begin{picture}(14.,15.)(0.,-4.) \thicklines
\put(-.2,0){\vector(0,1){9.5}}
\thinlines
\multiput(-.35,1.)(0,1.0){8}{\line(1,0){0.15}}
\put(-1.2,2.){200}
\put(-1.2,4.){400}
\put(-1.2,6.){600}
\put(-1.3,8.5){MeV}
\multiput(0,0)(.3,0.){3}{$\, _\times$}
\multiput(1.5,1.77)(.3,0.){3}{$\, _\times$}
\multiput(3.0,2.54)(.3,0.){3}{$\, _\times$}
\multiput(4.5,3.79)(.3,0.){3}{$\, _\times$}
\multiput(6.5,2.97)(.3,0.){3}{$\, _\times$}
\multiput(8.0,4.46)(.3,0.){3}{$\, _\times$}
\multiput(9.5,5.92)(.3,0.){3}{$\, _\times$}
\multiput(11.0,7.33)(.3,0.){3}{$\, _\times$}

\multiput(0,0)(.3,.0){4}{\line(1,0){.1}}
\multiput(1.5,1.76)(.3,.0){4}{\line(1,0){.1}}
\multiput(3.0,2.87)(.3,.0){4}{\line(1,0){.1}}
\multiput(4.5,3.82)(.3,.0){4}{\line(1,0){.1}}
\multiput(6.5,3.00)(.3,.0){4}{\line(1,0){.1}}
\multiput(8.0,4.75)(.3,.0){4}{\line(1,0){.1}}
\multiput(9.5,6.14)(.3,.0){4}{\line(1,0){.1}}
\multiput(11.0,7.25)(.3,.0){4}{\line(1,0){.1}}

\multiput(0.3,0)(.15,.0){4}{\circle{.03}}
\multiput(1.8,1.45)(.15,.0){4}{\circle{.03}}
\multiput(3.3,2.50)(.15,.0){4}{\circle{.03}}
\multiput(4.8,3.50)(.15,.0){4}{\circle{.03}}
\multiput(6.8,3.17)(.15,.0){4}{\circle{.03}}
\multiput(8.3,4.38)(.15,.0){4}{\circle{.03}}
\multiput(9.8,5.55)(.15,.0){4}{\circle{.03}}
\multiput(11.3,6.65)(.15,.0){4}{\circle{.03}}

\put(0.5,-1.5){$N$}
\put(2.0,-1.5){$\Lambda$}
\put(3.5,-1.5){$\Sigma$}
\put(5.0,-1.5){$\Xi$}
\put(7.0,-1.5){$\Delta$}
\put(8.5,-1.5){$\Sigma ^*$}
\put(10.0,-1.5){$\Xi ^*$}
\put(11.5,-1.5){$\Omega$}
\end{picture} \end{center}

\newpage
\noindent Figure 2\\
Direct and exchange matrix elements for non-identical
baryons.

\begin{center} \setlength{\unitlength}{.8cm}
\begin{picture}(16.5,15.)(1.5,3.) \thicklines
\put(1.5,3){\line(1,2){1.5}}
\put(3,7){\line(-1,2){1.5}}
\put(3,6){\line(0,1){1}}
\put(3,6){\line(1,0){1.5}}
\put(4.5,6){\line(0,1){1}}
\put(3,7){\line(1,0){1.5}}
\put(3.1,6.4){$D(A)$}
\put(1.9,3.2){$N$}
\put(1.9,9.8){$N'$}

\put(4.7,6.5){\circle*{.1}}

\put(8.,3){\line(-1,2){1.5}}
\put(6.5,7){\line(1,2){1.5}}
\put(6.5,6){\line(0,1){1}}
\put(6.5,6){\line(-1,0){1.5}}
\put(5.,6){\line(0,1){1}}
\put(6.5,7){\line(-1,0){1.5}}
\put(5.1,6.4){$D(B)$}
\put(7.,3.2){$Y$}
\put(7.,9.8){$Y'$}

\put(11.5,3){\line(1,2){1.5}}
\put(13,7){\line(-1,2){1.5}}
\put(13,6){\line(0,1){1}}
\put(13,6){\line(1,0){1.5}}
\put(14.5,6){\line(0,1){1}}
\put(13,7){\line(1,0){1.5}}
\put(13.1,6.4){$D(A)$}
\put(11.9,3.2){$N$}
\put(11.9,9.8){$Y'$}

\put(14.7,6.5){\circle*{.1}}

\put(18.,3){\line(-1,2){1.5}}
\put(16.5,7){\line(1,2){1.5}}
\put(16.5,6){\line(0,1){1}}
\put(16.5,6){\line(-1,0){1.5}}
\put(15.,6){\line(0,1){1}}
\put(16.5,7){\line(-1,0){1.5}}
\put(15.1,6.4){$D(B)$}
\put(17.,3.2){$Y$}
\put(17.,9.8){$N'$}

\end{picture} \end{center}


\begin{references}
\bibitem {Nijm92}
                 Th. A. Rijken, P. M. M. Maessen and J. J. de Swart,
                 Nucl. Phys. {\bf A547}, 245c (1992), and further
                 references in this issue which covers the Shimoda
                 Symposium on Hypernuclear and Strange Particle
                 Physics.
\bibitem {HS86} G. Holzwarth and B. Schwesinger, Rep. Progr. Phys.
                {\bf 49}, 825 (1986).
\bibitem {MZ86} U.-G. Meissner and I. Zahed, in {\it Advances in
                Nuclear Physics}, edited by J. W. Negele and E. Vogt
                (Plenum Press, New York, 1986), Vol. 17, 143.
\bibitem {Russians} V. G. Makhankov, Y. P. Rybakov and V. I. Sanyuk,
               {\it The Skyrme Model -- Fundamentals, Methods,
                Applications}, Springer Series in Nuclear and Particle
                Physics, (Springer-Verlag, Berlin, 1993).
\bibitem {WW92}
                 T. S. Walhout and J. Wambach, Int. J. Mod. Phys. E
                 {\bf 1}, 665 (1992).
\bibitem {SSG93}
                 F. G. Scholtz, B. Schwesinger and H. B. Geyer,
                 Nucl. Phys. {\bf A561}, 542 (1993).
\bibitem {KE90}
                 G. K\"albermann and J. M. Eisenberg,
                 Phys. Lett. B {\bf 235}, 6 (1990).
\bibitem {KE92}
                 G. K\"albermann and J. M. Eisenberg,
                 Phys. Rev. D {\bf 46}, 446 (1992).
\bibitem {DMG+91a} 
                 C. B. Dover, D. J. Milliner, A. Gal and D. H. Davis,
                 Phys. Rev. C {\bf 44}, 1905 (1991).
\bibitem {DG84}
                 C. B. Dover and A. Gal,
                 in {\it Progress in Particle and Nuclear Physics},
                 edited by D. Wilkinson (Pergamon Press,
                 Oxford, 1984) Vol. 12, 171-239.
\bibitem {DMG+91b}
                 C. B. Dover, D. J. Milliner and A. Gal,
                 Phys. Rep.  {\bf 184}, 1 (1989).
\bibitem {MGD+85}  
                 D. J. Milliner, A. Gal, C. B. Dover and R. H. Dalitz,
                 Phys. Rev. C {\bf 31}, 499 (1985).
\bibitem {VM85}
                 R. Vinh Mau, M. Lacombe, B. Loiseau, W. N.
                 Cottingham and P. Lisboa, Phys. Lett. B {\bf 150},
                 259 (1985).
\bibitem {Kal94}
                 G. K\"albermann (private communication).
\bibitem {JJPSW89}
                   P. Jain, R. Johnson, N. W. Park, J. Schechter and
                   H. Weigel, Phys. Rev. D {\bf 40}, 855 (1989).
\bibitem {PSW91}
                 G. Pari, B. Schwesinger and H. Walliser, Phys. Lett.
                 B {\bf 255}, 1 (1991).
\bibitem {GL85}
                 J. Gasser and H. Leutwyler, Nucl. Phys. {\bf B250}, 465
                 (1985).
\bibitem {YA88}
                 H. Yabu and K. Ando, Nucl. Phys. {\bf B301}, 601
                 (1988).
\bibitem {WSP90}
                 H. Weigel, J. Schechter and N. W. Park, Phys. Rev. D
                 {\bf 42}, 3177 (1990).
\bibitem {SRA1}
                 B. Schwesinger and H. Weigel, Phys. Lett. B {\bf
                 267}, 438 (1991).
\bibitem {SRA2}
                 B. Schwesinger and H. Weigel, Nucl. Phys. {\bf A450},
                 461, (1992).
\bibitem {YA85}
                 H. Yabu and K. Ando, Prog. Theor. Phys. {\bf 74},
                 750 (1985).
\bibitem {JJP85}
                 A. Jackson, A. D. Jackson and V. Pasquier, Nucl. Phys.
                 {\bf A432}, 567 (1985).
\bibitem{YSH89}
                 H. Yabu, B. Schwesinger and G. Holzwarth, Phys. Lett.
                 B {\bf 224}, 25 (1989).
\bibitem {Wal91}
                 H. Walliser, Nucl. Phys. {\bf A524}, 706 (1991).
\bibitem {S92}
                 B. Schwesinger, Nucl. Phys. {\bf A537}, 253 (1992).
\bibitem{Sch86}
                 J. Schechter, Phys. Rev. {\bf D34}, 868 (1986).
\bibitem{GJJS86}
                 H. Gomm, P. Jain, R. Johnson and J. Schechter, Phys.
                 Rev. D {\bf 33}, 3476 (1986).
\bibitem{KSS90}
                 V. B. Kopeliovich, B. Schwesinger and B. Stern, Phys.
                 Lett. B {\bf 242}, 145 (1990).
\bibitem{KS87}
                 V. B. Kopeliovich and B. E. Stern, JETP Lett. {\bf
                 45}, 203 (1987).
\bibitem{VWWW86}
                 J. J. M. Verbarschot, T. S. Walhout, J. Wambach
                 and H. W. Wyld, Nucl. Phys. {\bf A461}, 603 (1986).
\bibitem {BLRS85}
                  A. P. Balachandran, F. Lizzi, V. G. Rodgers and A. Stern,
                  Nucl. Phys. {\bf B256}, 525 (1985).
\bibitem {M93}
                 B. Moussalam, Ann. Phys. (N.Y.) {\bf 225}, 264 (1993).
\bibitem {DeSwart}
                  J. J. de Swart, Rev. Mod. Phys. {\bf 35}, 916
(1963). \end{references}
\end{document}